\documentclass[11pt,a4paper]{article}
\usepackage{graphicx}

\newtheorem{axiom}{Theorem}[section]
\newtheorem{guess}{Proposition}[section]

\newtheorem{definition}{Definition}[section]
\newtheorem{remark}{Remarks}[section]

\begin{document}

\bibliographystyle{unsrt}

\def\boxit#1#2{\setbox1=\hbox{\kern#1{#2}\kern#1}%
\dimen1=\ht1 \advance\dimen1 by #1 \dimen2=\dp1 \advance\dimen2 by
#1
\setbox1=\hbox{\vrule height\dimen1 depth\dimen2\box1\vrule}%
\setbox1=\vbox{\hrule\box1\hrule}%
\advance\dimen1 by .4pt \ht1=\dimen1 \advance\dimen2 by .4pt
\dp1=\dimen2 \box1\relax}

\def\build#1_#2^#3{\mathrel{\mathop{\kern 0pt#1}\limits_{#2}^{#3}}}

\def\K{\Bbb K}
\def\C{\Bbb C}
\def\R{\Bbb R}
\def\N{\Bbb N}
\def\ecart{\noalign{\medskip}}
\font\tenmsb=msbm10 \font\sevenmsb=msbm7 \font\fivemsb=msbm5
\newfam\msbfam
\textfont\msbfam=\tenmsb \scriptfont\msbfam=\sevenmsb
\scriptscriptfont\msbfam=\fivemsb
\def\Bbb#1{{\fam\msbfam\relax#1}}

\begin{titlepage}
\title{ Numerical study of the local contractivity\\ of the $\Phi_0^4$
mapping}
\author{by\\
Marietta Manolessou\\
and \\
Sofiane Tafat\\
\ \ \\
 E.I.S.T.I.\\
    Avenue du Parc   95011   Cergy-Pontoise-Cedex}
               \end{titlepage}

\maketitle

\begin{abstract}

Previous results on the non trivial solution of the
$\Phi^4$-equations of motion for the Green's functions
 in the Euclidean\  space (of  $0\leq r\leq 4$ dimensions)
 in the Wightman Quantum Field theory framework, are reviewed
in the $0-$ dimensional case from the following two aspects:

\begin{itemize}
\item{}(cf.\cite{MM})\
The structure of the subset $\Phi \subset {\cal B}$
 characterized by the bounds signs and ``splitting'' (factorization properties)
 is reffined and more explictly described in terms of a new closed subset $\Phi_0 \subset \Phi \subset {\cal B}$.
 Using a new norm we establish the  local contractivity of the corresponding $\Phi ^4_0$ mapping
in the neighborhood of a nontrivial sequence $H_0\in  \Phi_0$.
\item{} A new $\Phi ^4_0$  iteration is defined in the neighborhood of the sequence $H_0\in  \Phi_0 $.

In this paper we present  the results of our
numerical study, so:
\begin{itemize}
\item{a)} \emph{the stability of $\Phi_0$} \it{i.e.} the splitting,  bounds and sign
properties is clearly illustrated in the neighborhood of $H_0\in  \Phi_0$.
\item{b)}
the \emph{rapid convergence of this iteration} to the fixed point
is perfectly realized thanks to the new starting points of the iteration.
\end{itemize}
\end{itemize}
\end{abstract}

\vfill\eject

\listoffigures
\vfill\eject

\pagenumbering{arabic}

\section{ Introduction}
\subsection{A new non perturbative method -
the new recent results}

\noindent\ \   Several years ago we started a program for
 the construction of a non
 trivial $\Phi^4_4$  model consistent with the
general principles of a Wightman Quantum Field
 Theory
($Q.F.T.$) \cite{(Q.F.T.)}.
In reference \cite{MM1} we have
introduced
 a non perturbative method for the
 construction of a non trivial solution of
 the system
 of the $\Phi^4$ equations of motion
for the Green's functions, in the Euclidean space
 of zero, one and two dimensions.
 In reference  \cite{MM2}
we tried to apply an extension of this method to the case
 of four-(and a fortiori of three-)dimensional
Euclidean momentum space.

The general aspects of the method together with its comparison and validity arguments with respect to other non perturbative methods are presented in these previous references, and in particular in the theoretical aspect of this last study \cite{MM}.

In this paper we present the numerical study of a new  $\Phi^4_0 $-Iteration.

We use  the zero dimensional analog of the system of
equations of motion  introduced in the previous papers.
  The reasons that motivated us for a study in smaller dimensions and
 not directly in four, were
 the absence of the difficulties due to the renormalization and the pure
 combinatorial character of the problem in zero dimensions.
 Another useful aspect  of the zero dimensional case is the fact that
 it provides a direct way to
test numerically the validity of the method.

  What  are the new developments in the present zero dimensional study:
  \begin{enumerate}
    \item{} (cf.\cite{MM})The ``new $\Phi_0$ subset is explicitly given in terms of the
    "splitting" sequences upper $\left\{ \delta_{n, max} (\Lambda) \right\}$ and
    lower $\left\{ \delta_{n, min}(\Lambda) \right\}$ envelops.
    \item{} (cf.\cite{MM}) The non triviality and stability of the subset $\Phi_0$ under the mapping ${\cal M^*}$ is established
    in terms of the new basic sequences $\left\{ \delta_{n, max} (\Lambda) \right\}$,
   $\left\{ \delta_{n, min}(\Lambda) \right\}$ and $\left\{ H_{0} \right\}$ The latter is furthermore used
    for the proof of  local contractivity.
    This proof is simpler in comparison with our previous
    analogs,  due also to the fact that we introduce a new norm on the Banach space $\mathcal{B}$.

    \item{} In the present paper, starting from these particular sequences, $\left\{ \delta_{n, max} (\Lambda) \right\}$ and
    $\left\{ \delta_{n, min}(\Lambda) \right\}$, we define a
    new $\Phi_0$-iteration and explore the behavior of the Green's functions
    (essentially the $\delta$ functions), at sufficiently large $n$
    and reasonable order of this new iteration.

   These last numerical  results are convincing, the convergence is rapidly established for
    different values of $\Lambda$ and the sign and  splitting
    properties (\emph{stability+contractivity}) give coherent results with respect to our
    theoretical conclusions.

  \end{enumerate}

\vfill\eject

\subsection{Reminders}\
In \cite{MM} we presented in detail the definitions and results introduced in the previous papers together with the  new ones. Let us present  only the necessary among them, for the best understanding of our  numerical study.
\subsubsection{The $\Phi^4_0$\,
 equations of motion  the subsets $\Phi_0\subset\Phi$ and the new mapping ${\cal M}^*$}
 
\begin{definition}[The $\Phi^4_0$\,
 equations of motion]\label{def.1.1}\ \

$\forall \ \Lambda\, \in\Bbb R^{+}$
  \begin{equation}  H^2 (\Lambda) =  - \Lambda H^4(\Lambda) \ \ +1
\end{equation}
 and for all $n \geq3$,
\begin{equation}
H^{n+1}(\Lambda) = \  \ A^{n+1}(\Lambda) +  B^{n+1}(\Lambda) +
C^{n+1}(\Lambda)
 \end{equation}

with:
\begin{equation}
A^{n+1}(\Lambda) =\,  - \Lambda H^{n+3}(\Lambda);
\end{equation}
\begin{equation}
B^{n+1}(\Lambda) =\,  - 3\Lambda\sum_{\varpi_n(J)}{n \ !\over
j_{1}!j_{2}!} H^{j_{2}+2}(\Lambda) H^{j_{1}+1}(\Lambda);
\label{1.1.6}
\end{equation}
\begin{equation}
C^{n+1}(\Lambda) =\, - 6\Lambda\sum_{\varpi_n(I)} {n\ !\over
i_{1}!i_{2}!i_{3}!\ \sigma_{sym}(I)} \prod_{l=1,2,3}H^{i_{l}+1}
(\Lambda) \label{1.1.7}
\end{equation}
\end{definition}
Here the notation $\varpi_n(J)$, means the set of different
partitions
 $(j_{1};j_{2})$ of $n$ such
that  $j_{1}$ is an odd integer and $j_{1}+j_{2}=n$.
 Respectively  $\varpi_n(I)$ is the set of
 triplets of odd numbers-different ordered partitions $(i_{1};i_{2};i_{3})$
 of $n$ with $ i_{1}\geq
i_{2}\geq i_{3}$.

The symmetry-integer $\sigma_{sym}(I)$ is defined by:
\begin{equation}
\sigma_{sym}(I) = \left\{\matrix{\hfill 3!& if&
i_{1}=i_{2}=i_{3}\cr
                        \hfill  1& if&   i_{1}\neq i_{2}\neq i_{3}\neq i_{1}\cr
                     \hfill&2& otherwise  \cr}\right\}
\end{equation}
\vspace{5mm}

\subsubsection{The vector space ${\cal B}$   }

\begin{definition}   \ \

 We introduce the vector space {${\cal B}$}
of the sequences $H=\{ H^{n+1}\} _{n =2k+1;  k\in\Bbb N}$
 by the following:
 The functions $H^{n+1}(\Lambda)$ belong to the space
 $C^{\infty}(\Bbb R^+ )$
 of continuously
differentiable numerical functions of the variable $\Lambda \in
\Bbb R^+$
 (which physically represents the coupling constant).

 Moreover, there exists a universal (independent
 of n and of $\Lambda$)
 positive constant $K_0$, such that the following uniform
 bounds are verified:

$$\forall  \ n=2k+1, k\in \Bbb N $$
\begin{equation}
|H^{n+1}(\Lambda)|\leq n\ !  (K_0)^n\qquad \forall\, \Lambda \in
\Bbb R^+
\end{equation}
\end{definition}

We suppose that the system of equations under consideration,
 concerns always
 (following our introduction and
the previous definition) the sequences of Euclidean connected
 and amputated with respect to the
free propagators Green's functions (the Schwinger functions). and
that these sequences  denoted by $H=\{ H^{n+1}\}_{n =2k+1\;,
k\in\Bbb N}$ belong to the above space   ${\cal B}_c$.

\subsubsection{The splitting sequences and the subsets $\Phi_0\subset\Phi\subset {\cal B}$}
\begin{definition} \ \

 We first introduce the class {${\cal D}$} of sequences
 $$\delta= \{\delta_{n}
 (\Lambda)\}
_{n=2k+1; k\in\Bbb N}\in {\cal B},$$
 such that they verify the bounds $(2.1.1)$ in the following simpler
 form:
\begin{equation}
|\delta_{n} (\Lambda)|\leq \,   K_0  ,
\end{equation}
$$   \forall\,  n=2k+1;\  k\in \Bbb N  $$
\end{definition}

\begin{definition}{splitting and signs-$\Phi \subset{\cal B}$}\label{def.1.3}\ \

\emph{ We shall say that a sequence $H \in {\cal B}$ belongs to
the subset {$\Phi \subset{\cal B}$}
 if there exists an increased associated sequence
 of positive and bounded functions on
 $\Bbb R^+$,
$$\delta= \{\delta_{n} (\Lambda)\}
_{n=2k+1; k\in\Bbb N}\in {\cal D},$$
 such that  the following
 ``splitting'' (or factorization) and sign
 properties are verified:$$\forall \Lambda \in \Bbb R^+$$}

\begin{description}
\item[$\Phi.1$]
 \begin{equation}
  H^2 (\Lambda) =  1  +\Lambda \delta_1 (\Lambda)\,  \hbox{ with:}\,
  \,   \build\lim_{\Lambda \rightarrow 0}^{}
\delta_1 (\Lambda) = 0 \label{1.2.10}
\end{equation}

\item[$\Phi.2$]
 \begin{equation}   H^4 (\Lambda) =  -  \delta_3 (\Lambda)
\lbrack     H^2 (\Lambda)\rbrack^3,\,    \hbox{ with:}\,  \,
  \delta_3 (\Lambda)\leq 6\Lambda ,\   \build\lim_{\Lambda \rightarrow 0}^{}
{\delta_3 (\Lambda)\over \Lambda} = 6 \label{1.2.11}
\end{equation}

\item[$\Phi.3$]
\begin{equation}
\forall n\geq 5 \qquad
    H^{n+1} (\Lambda) = {\displaystyle  \delta_n (\Lambda)  C^{n+1}\over \displaystyle
 3\Lambda n(n-1)}
 ,  \  \hbox{ with :}\,  \  \build\lim_{\Lambda \rightarrow 0}^{}
 {\displaystyle\delta_n (\Lambda)\over
\displaystyle \Lambda} =  3n(n-1) \label{1.2.12}
\end{equation}

\item[$\Phi.4$]\ \

 \emph{$\forall\, \, n=2k+1$\, with\,  $k\in\Bbb N^*,\, \exists$
  positive
 continuous functions of $\Lambda$,\, $$ \delta_{n,max}(\Lambda),\
 \delta_{n,min}(\Lambda)$$
(uniform bounds independent on $H$), $\delta_{max}^{\infty}$,
$\delta_{\infty}$
 (uniform limit and bound at infinity),
 such that:}
 \begin{equation}
\delta_{n,max}(\Lambda) \  >\  \delta_{n,min}(\Lambda)
\label{1.2.13}
\end{equation}
\begin{equation}
\delta_{n,min}(\Lambda)\leq \delta_{n}(\Lambda)
\leq\delta_{n,max}(\Lambda) \label{1.2.14}
\end{equation}
\begin{equation}
\build\lim_{n \rightarrow \infty}^{}\delta_{n,max} \leq
\delta_{max}^{\infty} (\Lambda)\leq \delta_{\infty}\label{1.2.15}
\end{equation}
\end{description}
\end{definition}

\begin{definition}
\begin{equation}
    \delta_{3, max}(\Lambda) =6\Lambda;\qquad
 \delta_{3, min}(\Lambda) = {6\Lambda\over 1+9\Lambda(1+6\Lambda^2)}
\label{1.2.17}
 \end{equation}
\emph{and}\,  $\forall\, \,  n\geq 5 $
 \begin{equation}\delta_{n, max}(\Lambda)\, = {3\Lambda\ n(n-1)\over 1+ 3\Lambda\ n(n-1)d_0 }
\end{equation}
\emph{here we put} $d_0=0.001$
\begin{equation}\delta_{n, min}(\Lambda)\, = {3\Lambda\ n(n-1)\over 1+3\Lambda\
 n\ (n-1) }
\end{equation}
\label{def. 2.4}
\end{definition}

\begin{definition} \ \

 \emph{By using  $\{\delta_{n, max} (\Lambda)\} _{n=2k+1;
k\in\Bbb N}$\, and\,
 $\{\delta_{n, min} (\Lambda)\}
_{n=2k+1; k\in\Bbb N}$
 introduced before we define the following
 sequences:}
$$\{H^{n+1}_{max} \}
_{n=2k+1; k\in\Bbb N}\in {\cal B}\,  \hbox{\emph{and}}\,
 \{  H^{n+1}_{min}\}
_{n=2k+1; k\in\Bbb N}\, \in {\cal B}$$

\begin{equation}
  H^2_{max}(\Lambda )=(1+6\Lambda^2)^2;\
  \qquad   H^2_{min} =1
\end{equation}
\begin{equation}
  H^4_{max}(\Lambda )=- 6\Lambda  [H^2_{max}]^3;\ \
H^4_{min}
(\Lambda )= -\delta_{3, min}(\Lambda)
\end{equation}

 \emph{ and recurrently   for every $n\geq 5$ :}
\begin{equation}
  H^{n+1}_{max}= {\delta_{n,max} (\Lambda) C^{n+1}_{max}\over 3\Lambda\
n(n-1)}
\end{equation}
\begin{equation}
  H^{n+1}_{min}(\Lambda)= {\delta_{n,min}(\Lambda)  C^{n+1}_{min} \over
3\Lambda\ n(n-1)}
\end{equation}
\label{def.2.5}
\end{definition}

\begin{definition}\ \textbf{The subset} $\Phi _0$\label{def.2.7}

Taking into account the sequences  of the previous definition  \  we introduce
 the following subset $\Phi_0\subset \Phi$:
\begin{equation}
\Phi _0=\left\{ H\in\Phi : \vert H^{n+1}_{min}\vert\ \leq
\vert H^{n+1}\vert\leq
\vert H^{n+1}_{max}\vert , \quad  \forall n=2k+1,\ k\in\ \N\right  \}
\label{3.60}
\end{equation}
\end{definition}

Using the previous sequences we defined in \cite{MM}  the \emph{``fundamental sequence''.}

\begin{definition}\label{def.2.5}
 \begin{equation}
    H^2_0 (\Lambda) =  1-\Lambda H^4 _{min}
    \ \ \
    \end{equation}
    \begin{equation}
   H^{4}_{0} (\Lambda) = - \delta_{3, 0}(\Lambda)[H^{2}_{0}]^3\quad
\hbox{with}\quad \delta_{3, 0} (\Lambda)= \displaystyle{ \frac{6\Lambda}{1+9\Lambda
 - \displaystyle{\frac{\Lambda
|H^6_{min}|}{ |H^4_{min}|} }}}
\label{1.2.52}
\end{equation}
and for every $n\geq 5$
    \begin{equation}
H^{n+1}_{0}(\Lambda) = {\delta_{n, 0}(\Lambda)
 C^{n+1 }_{0}( \Lambda) \over 3\Lambda
 n (n-1)};
\label{1.2.53}
\end{equation}
       with:
\begin{equation}C^{n+1}_0(\Lambda) = - 6\Lambda\sum_{\varpi_n(I)}
 {n\ !\over i_{1}!i_{2}!i_{3}!\
 \sigma_{sym}(I)}
\prod_{l=1,2,3}H^{i_{l}+1}_0 (\Lambda);
\end{equation}

\begin{equation}
\delta_{n,0}(\Lambda)=\frac{3\Lambda n(n-1)}
 {1+D_n(H_{(min) })}
\label{1.2.54}
\end{equation}
and
 \begin{equation}
D_n(H_{(min)})=\displaystyle{ \frac{|B^{n+1}_{min}| - |A^{n+1}_{min}| }{|H^{n+1}_{max}|}}
 \label{1.2.55}
\end{equation}
\end{definition}
and proved,
\begin{guess}{(the non triviality)}

The set of sequences  {$\Phi$} given by the
definition  \ref{def.1.3} is a nontrivial subset of the space
 ${\cal B}$.
\label{prop.2.1}
\end{guess}

\vspace{3mm}
Then, we introduced
   \begin{definition}\, {The new mapping ${\cal M^*}$}\ \label{def.1.6}\

 $${\cal M^*}:\, \Phi \stackrel{\cal M^*}
\longrightarrow
 {\cal B}$$

\begin{equation}
H^{2'} (\Lambda) = 1+\Lambda \delta_1^{'}(\Lambda)\qquad
\hbox{with}\qquad \delta_1^{'}(\Lambda)= -H^4 (\Lambda)
\label{1.2.51}
\end{equation}
\begin{equation}
H^{4'} (\Lambda) = - \delta_3'(\Lambda)[H^{2'}]^3\quad
\hbox{with}\quad \delta_3^{'}(\Lambda)= 6\Lambda[1+6\Lambda
H^2(3/2
 - \frac{|H^6|}{6|H^4||H^2|})]^{-1}
\label{1.2.52}
\end{equation}
and for every $n\geq 5$
    \begin{equation}
H^{n+1'}(\Lambda) = {\delta_n^{'}(\Lambda)
 C^{n+1'}( \Lambda) \over 3\Lambda
 n (n-1)};
\label{1.2.53}
\end{equation}
    and:
\begin{equation}\delta_n^{'}(\Lambda)=\frac{3\Lambda n(n-1)}
 {1+D_n(H)}
\label{1.2.54}
\end{equation}
 with
 \begin{equation}
D_n(H)={|B^{n+1}| - |A^{n+1}| \over |H^{n+1}|} \label{1.2.55}
\end{equation}
\end{definition}

 \vspace{2mm}
and proved:
\begin{axiom}
{The stability of the subset $\Phi_0$}\ \ \label{Th.1.2.}

If $H\in \Phi$; then ${\cal M^*}(H)\subset \Phi$
  under the condition:
$$0<\Lambda\leq 0.05$$
\label{Th.1.2}
\end{axiom}

Furthermore (always in \cite{MM}), we constructed a Banach space ${\cal B}_c  \subset {\cal B}$ by introducing the following norm ${\cal N}$: 
\begin{definition}\ \ \label{def.norm}\
\begin{equation}\begin{array}{l}
{\cal N}:\ {\cal B}\rightarrow \Bbb R^+\  \\
\ \ \ \ H\mapsto\|H\|\ \  \hbox{with:}\ \  \|H\| =\{ \displaystyle{
\sup_{\Lambda;n }}\displaystyle{\frac{|H^{n+1}|}{M_n}
\}}
\end{array}
\label{3.1.94}
\end{equation}
Here $\forall\ \  \Lambda\in\Bbb R^+$:
\begin{equation}
M_1 (\Lambda)=H^2_{max}=(1+6\Lambda^2)^2;\quad M_3 (\Lambda)=\,
 \delta_{3max}(M_1)^{3};\  \ \label{3.1.95}
\end{equation}
and for every $n\geq 5$
\begin{equation}\begin{array}{l}
M_n (\Lambda)=\ n(n-1) \delta_{nmax} M_{n-2}(M_1)^2\ \ \ \ \  
\end{array}
\label{3.1.96}
\end{equation}
\label{def.4.1}
\end{definition}
The  function ${\cal N}$ above
defines
 a finite
norm inside
 a non empty subspace ${\cal B}_c$ of ${\cal B}$.
 This subspace ${\cal B}_c$
obviously contains $\Phi_c$ and is a Banach space with
 respect to the ${\cal N}$- topology.

Using the above definition of the norm
 we introduced a ball-neighborhood $S_{\rho}(H_0)$  of the
{fundamental sequence} and  by theorem \ref{Th.3.1}, we have established the  local contractivity of ${\cal M^*}$
inside it by the following:
\vspace{3mm}

\begin{axiom}  {\textbf{the local
contractivity of the mapping
 ${\cal M^*}$\ \
in  $S_{\rho}(H_0)\ \subset \Phi_0$}} \label{Th.3.1} 

There exists a finite positive
constant
 $\Lambda^*(\approx 0.45)$
 such that the mapping ${\cal M^*}$ is locally contractive in $S_{\rho}(H_0)$ consequently there exiists  a unique non
trivial
 solution of the $\Phi_0^4$
 equations of motion in the neihbourhood $S_{\rho}(H_0)\ \subset
\Phi_0$  of the fundamental sequence
$H_0$.
\end{axiom}
\vfill\eject

\vfill\eject

\section{The numerical study}

\subsection{The different aspects of the analysis}

We have studied three different aspects of our study,
consequently we have obtained three sets of figures, that we
describe in detail  in what follows.

The general conclusion of this numerical experience appears clearly
the same in all  three sets.

We notice that the first three orders of the iteration of
$\delta_{n,max}$ and $\delta_{n,max}$ yield different curves which
 come closer   to each other  till the fourth order
iteration. Beyond, {\it i.e.} for $5^{th}$ and  $6^{th}$ order, we observe a  perfect coincidence of
$\delta_{n,max}$ and $\delta_{n,min}$.

 So, when the value of $\Lambda$ lies in $[0.001,\ 0.1]$, the
 neighborhood where lies the fixed point of the contractive mapping
 is manifestly around the $\{H_0\}$ sequence,
 (\emph{almost first order iteration of  $\{ \delta_{n,min}
 \}$ sequence}). This fact is enhanced by the following observation:

For a given value of $\Lambda$ we remark  that the sequence
$\{\delta_{n, max}\}$ decreases during the iteration procedure (resp
the sequence $\{\delta_{n, min}\}$ increases). The two sets are
almost the same up to the fourth iteration. We notice that the
decreasing rate of $\{\delta_{n, max}\}$ is more important than the
increasing rate of $\{\delta_{n, min}\}$, and this again underlines the
fact that the $\{H_0\}$ neighborhood is the best for the local
contractivity.

 This result is more satisfactory (from the point of view of the
 bound of $\Lambda$) in comparison with the theoretical proof of
 the validity of the contractivity criteriun at $\Lambda \leq 0.05$.

\subsubsection{First set of figures}
The first set of figures displays the convergence proof of the
mapping for different values of $\Lambda$, using as stating points
both $\delta_{n,max}$ and  $\delta_{n,min}$.
This set  represents the results of twenty
iterations of the mapping for different values of $\Lambda$\\ (i.e
$\Lambda \in \{ 0.001,\ 0.01,\ 0.03,\ 0.05,\ 0.075,\ 0.1 \}$) at $n$
fixed. We have chosen ten different values of $n$:
$$n=7,\ 9,\ 11,\ 13,\ 15,\ 17,\ 19,\ 21,\ 23,\ 25. $$

 The stability of the values is already attained at the tenth
 iteration for all values of $n$.

\subsubsection{Second set of figures}

The second set displays the convergence of the mapping up to the
sixth iteration, for different values of $\Lambda$.

\subsubsection{Third set}

The third set of figures  displays the summary  of the previous
configuration, for the sixth iteration.

This figure  represents the results of the mapping of $\delta_n$
functions,  for $$n=7,\ 9,\ 11,\ 13,\ 15,\ 17,\ 19,\ 21,\ 23,\ 25.$$
for different values of $\Lambda$\\
(i.e $\Lambda \in \{ 0.001,\ 0.01,\ 0.03,\ 0.05,\ 0.075,\ 0.1 \}$)
and for all   $20$ iterations.

The figure illustrates clearly the convergence of the iteration to
the fixed point. We remark also that the convergence is more rapid
for sufficiently small values of $\Lambda$ (and even for bigger
than the critical point $0.45$, due to the small values of $n$).

Our experience shows that the stability is impossible, for example
$n=1000$, when $\Lambda$ becomes bigger than $0.05$.

This figure  represents the results of the mapping of $\delta_n$
functions as surfaces of $n$ and $\nu$ for fixed $\Lambda$ (at the
six different values).

We remark in this figure that:
\begin{itemize}
    \item For small values of $\Lambda$ ($\Lambda \leq 0.001$),
    the "decrease" properties of $\delta_n$'s are not apparent.
    \item For the intermediate values ("good values") of
    $\Lambda$, the surfaces show the expected concavity as far as $\nu$ (iteration number)
    increases.
    \item For large values of $\Lambda$ (bigger than the critical value $\sim 0.045$), we observe a rapid
    increasing surface (because we are far from the stability and contractivity criteria).
\end{itemize}

\vfill\eject
\section{The figures}

\begin {enumerate}

\item{}

\begin{figure}[h]
\begin{center}
\hspace*{-5mm}
\includegraphics[width=6cm]{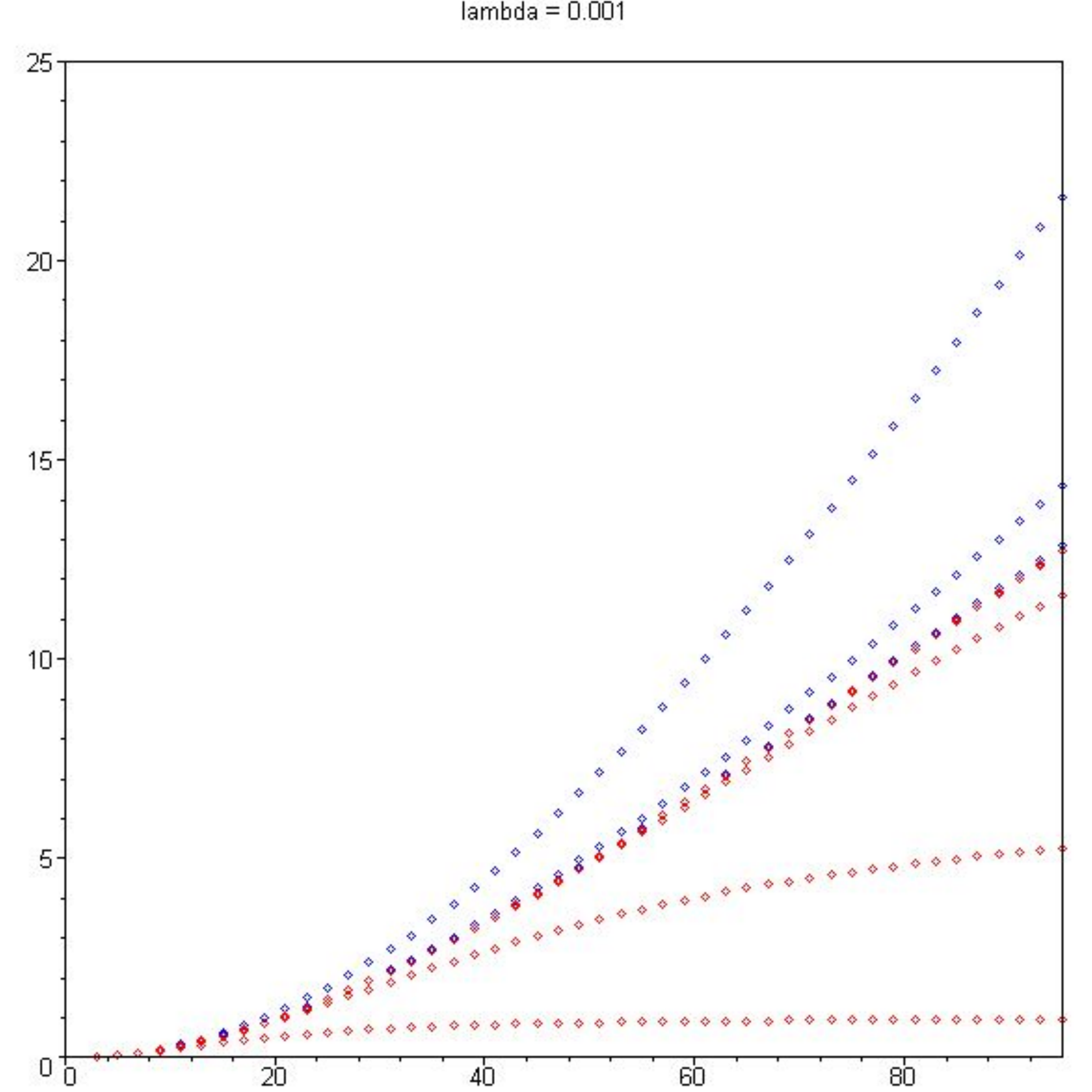}
 \caption{\underline{\textbf{$1^{rst}$ set:}}  Convergence up to $\nu= 6$ with
$\Lambda=0.001$  starting from $\delta_{n, max}$ and $\delta_{n, min}$}
 \label{fig.1}
 \end{center}
\end{figure}
\vspace{3mm}

\item{}

\begin{figure}[h]
\begin{center}
\hspace*{-5mm}
\includegraphics[width=6cm]{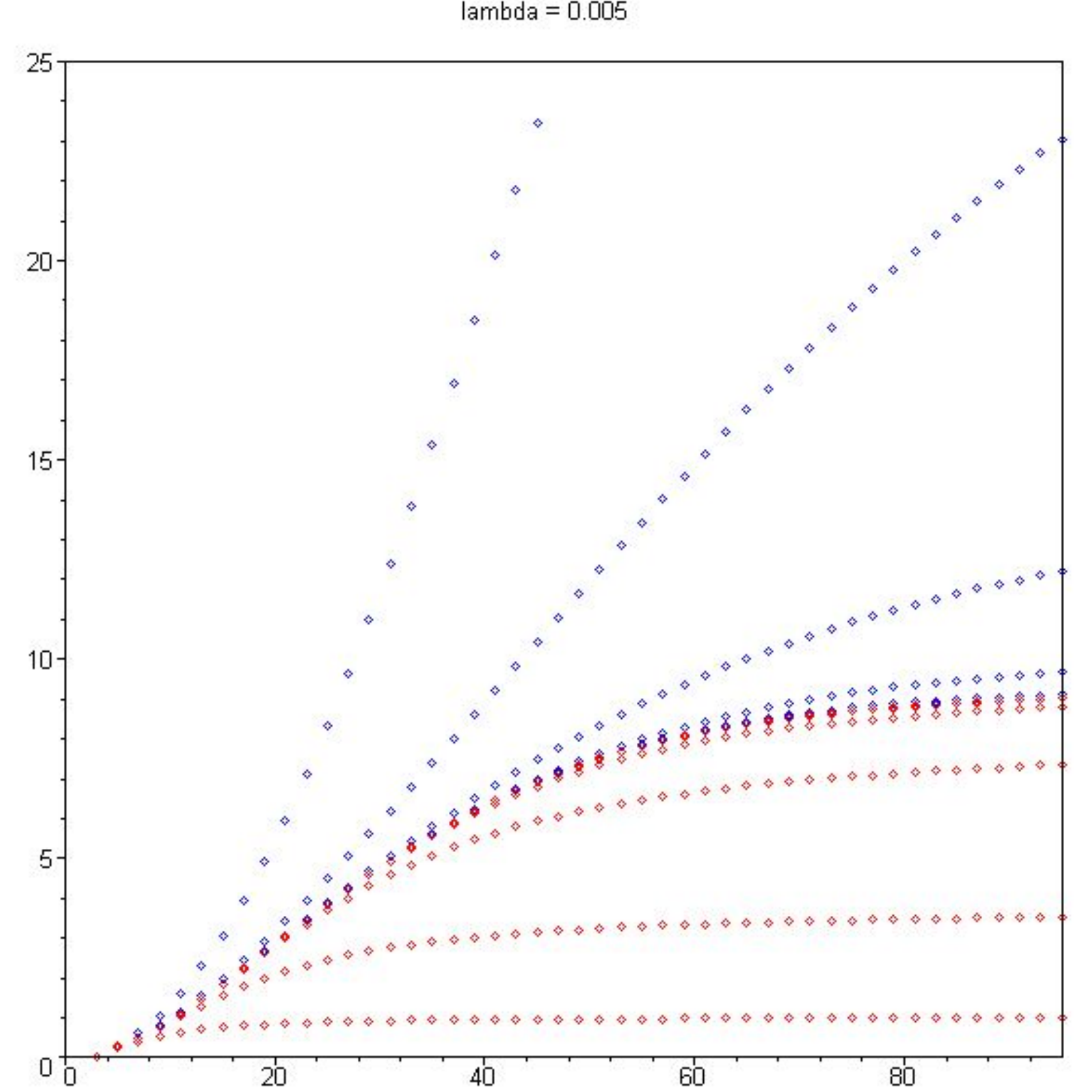}
 \caption{\underline{\textbf{$1^{rst}$ set:}}  Convergence up to $\nu= 6$ with
$\Lambda=0.0005$  starting from $\delta_{n, max}$ and $\delta_{n, min}$}
 \label{fig.2}
 \end{center}
\end{figure}
\vspace{3mm}

\item{}

\begin{figure}[h]
\begin{center}
\hspace*{-5mm}
\includegraphics[width=6cm]{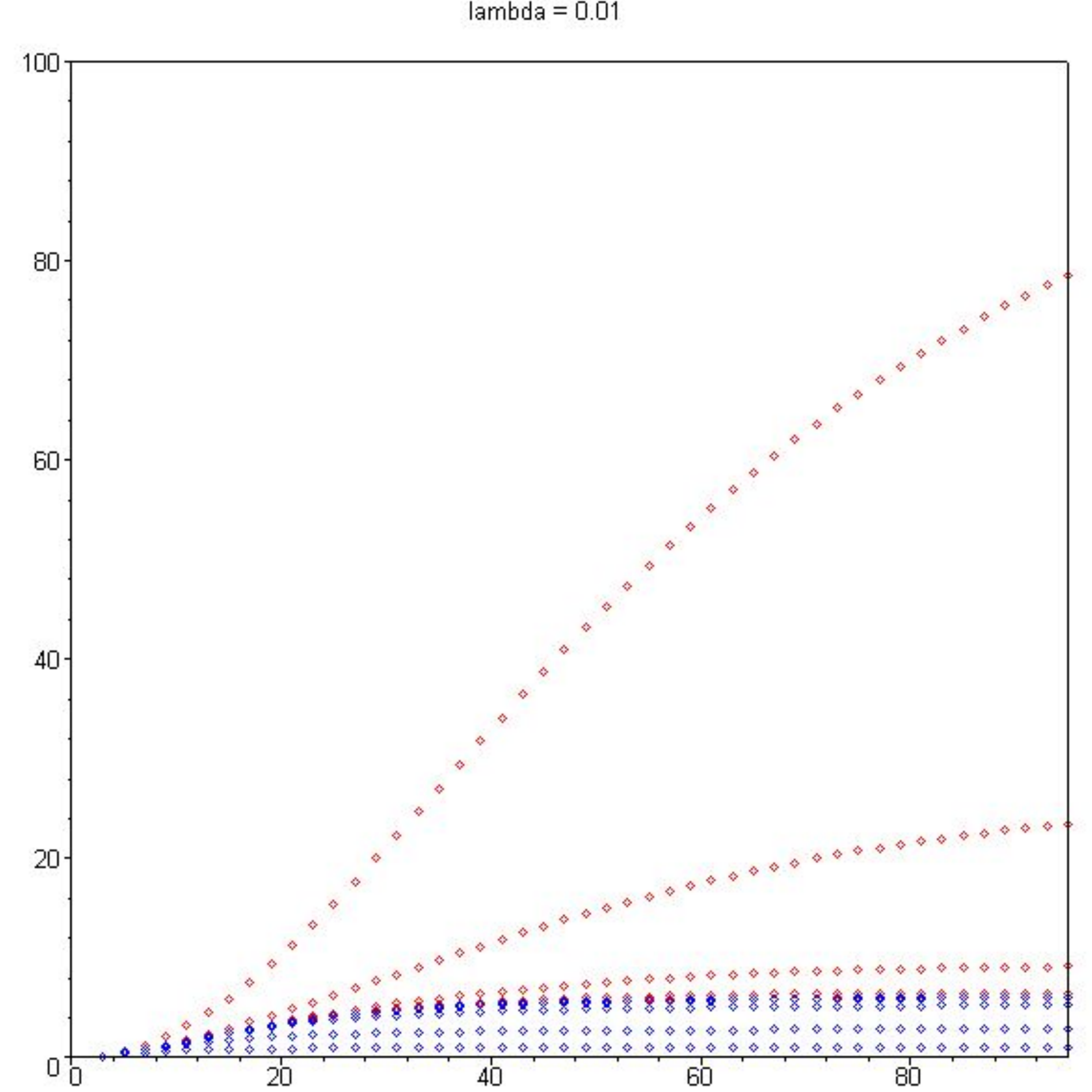}
 \caption{\underline{\textbf{$1^{rst}$ set:}}  Convergence up to $\nu= 6$ with
$\Lambda=0.01$  starting from $\delta_{n, max}$ and $\delta_{n,min}$}
 \label{fig.3}
 \end{center}
\end{figure}
\vspace{3mm}

\item{}

\begin{figure}[h]
\begin{center}
\hspace*{-5mm}
\includegraphics[width=6cm]{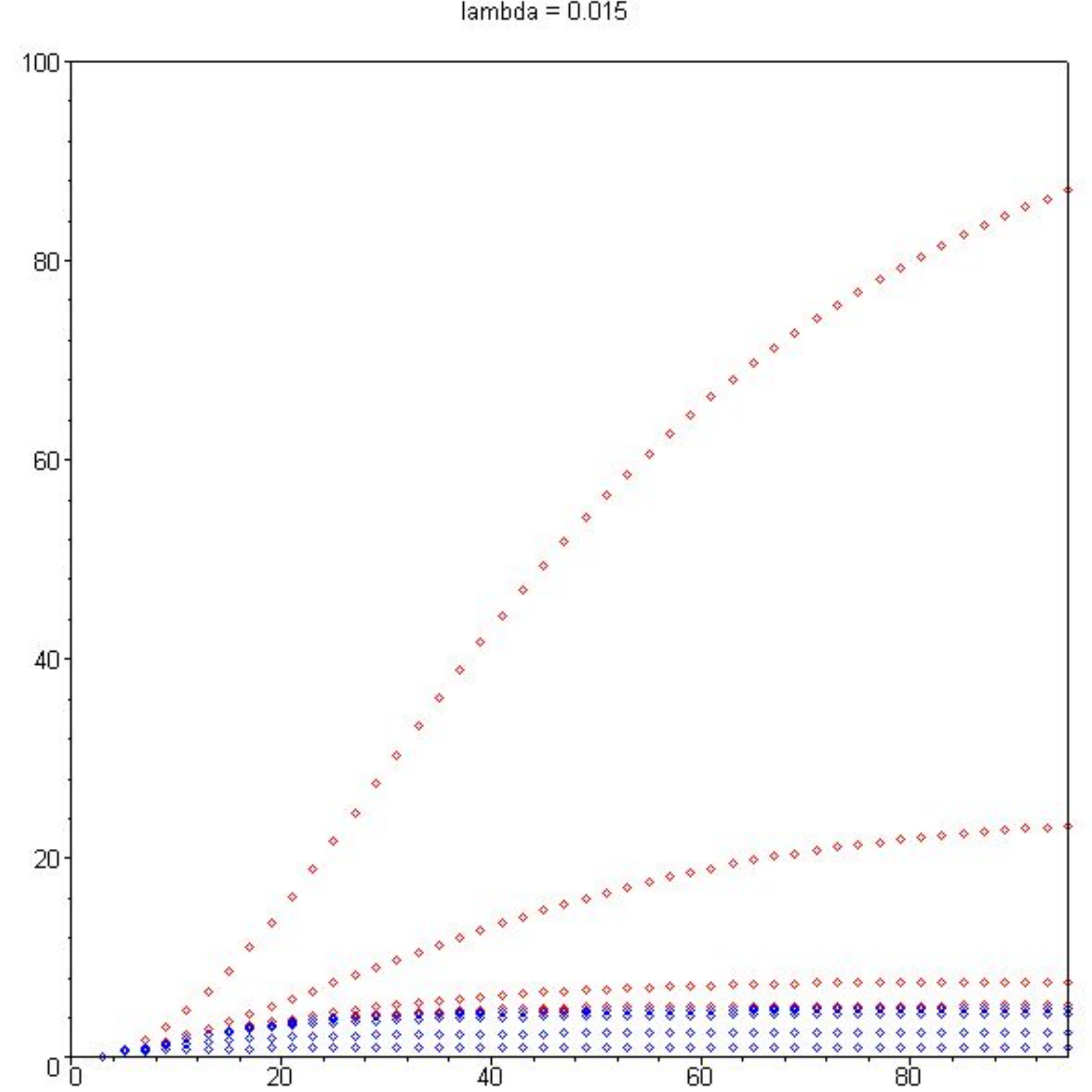}
 \caption{\underline{\textbf{$1^{rst}$ set:}}  Convergence up to $\nu= 6$ with
$\Lambda=0.015$  starting from $\delta_{n, max}$ and $\delta_{n, min}$}
 \label{fig.4}
 \end{center}
\end{figure}
\vspace{3mm}

\item{}
\begin{figure}[h]
\begin{center}
\hspace*{-5mm}
\includegraphics[width=6cm]{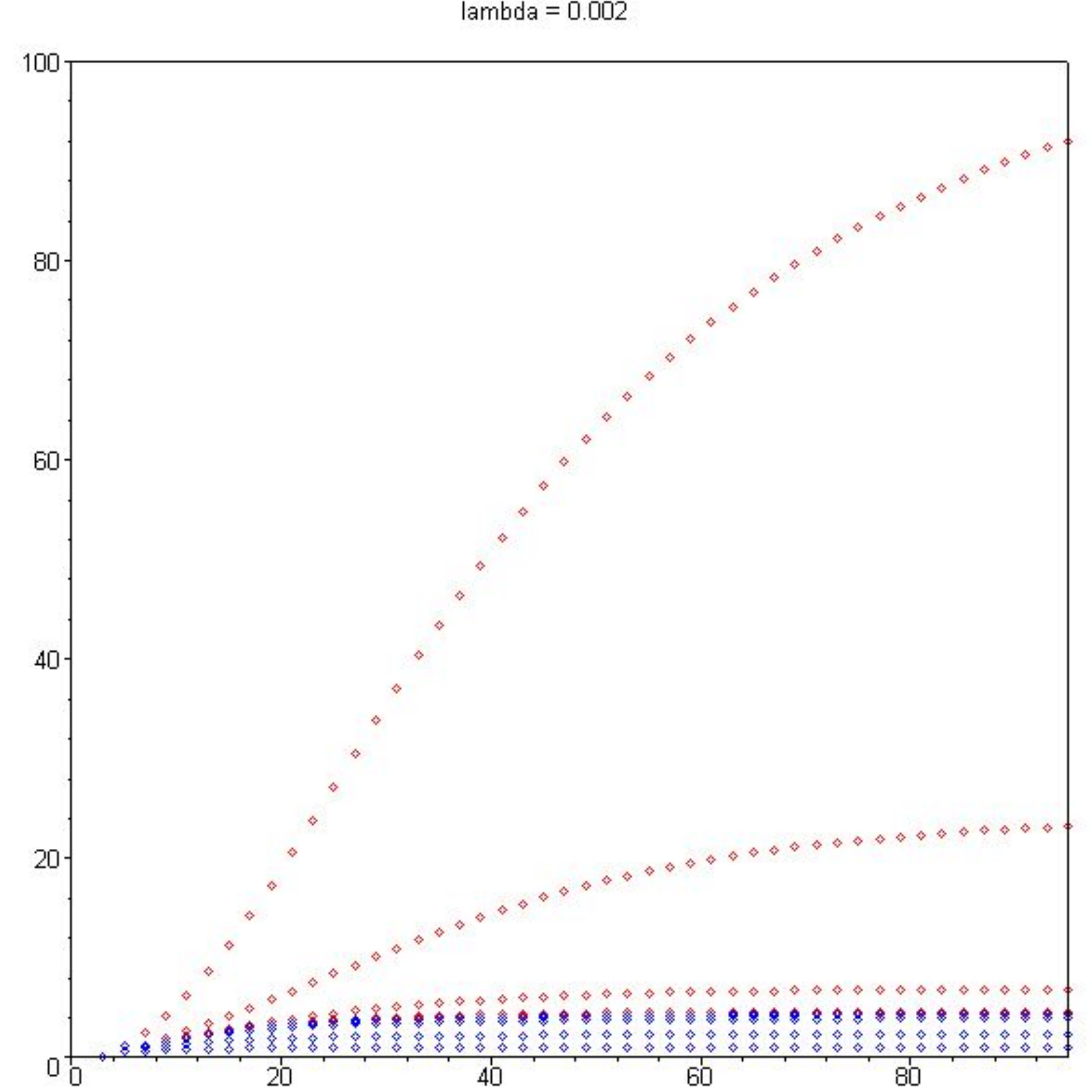}
 \caption{\underline{\textbf{$1^{rst}$ set:}}  Convergence up to $\nu= 6$ with
$\Lambda=0.02$  starting from $\delta_{n, max}$ and $\delta_{n, min}$}
 \label{fig.5}
 \end{center}
\end{figure}
\vspace{3mm}

\item{}

\begin{figure}[h]
\begin{center}
\hspace*{-5mm}
\includegraphics[width=6cm]{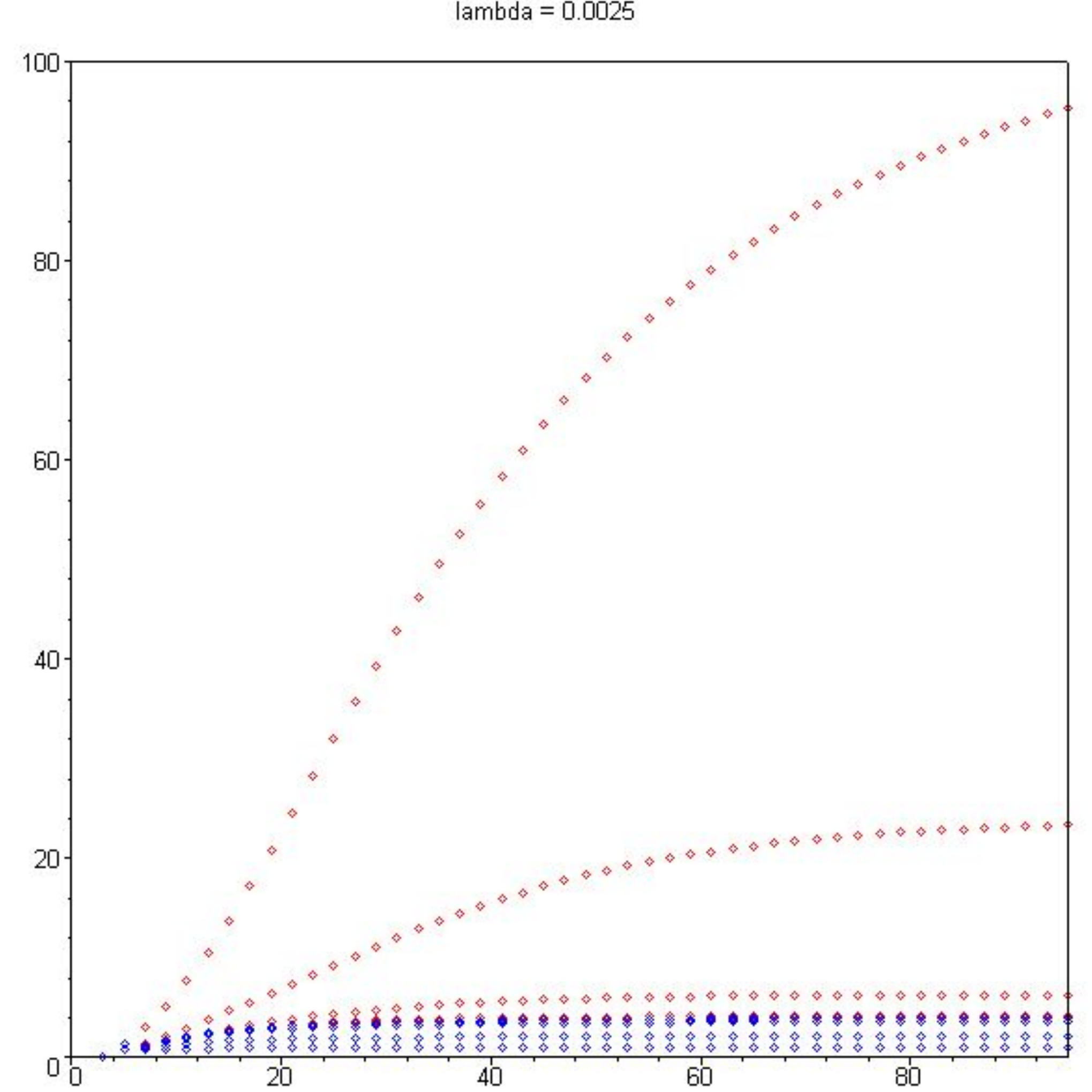}
 \caption{\underline{\textbf{$1^{rst}$ set:}}  Convergence up to $\nu= 6$ with
$\Lambda=0.025$  starting from $\delta_{n, max}$ and $\delta_{n, min}$}
 \label{fig.6}
 \end{center}
\end{figure}
\vspace{3mm}

\item{}

\begin{figure}[h]
\begin{center}
\hspace*{-5mm}
\includegraphics[width=6cm]{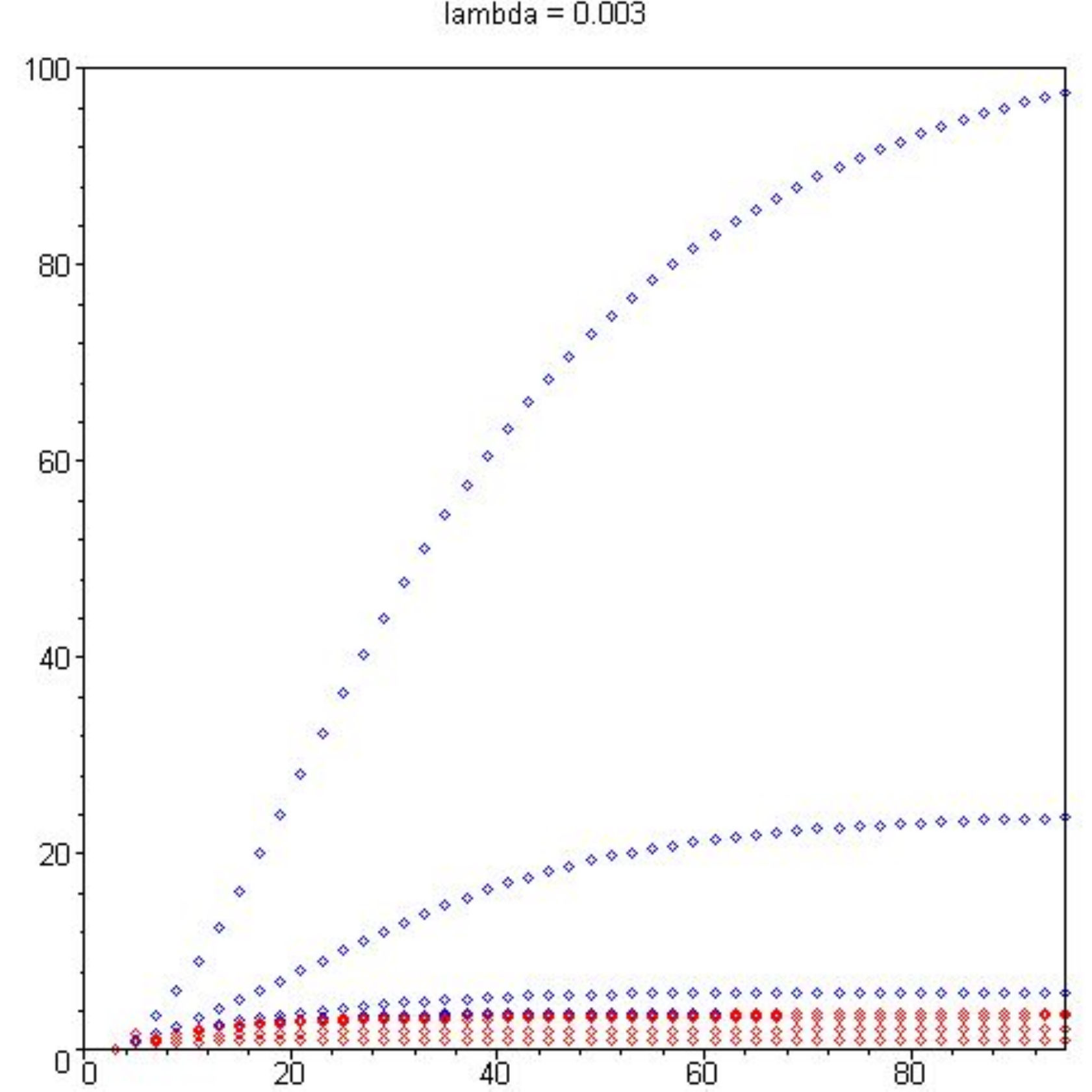}
 \caption{\underline{\textbf{$1^{rst}$ set:}}  Convergence up to $\nu= 6$ with
$\Lambda=0.03$  starting from $\delta_{n, max}$ and $\delta_{n, min}$}
 \label{fig.7}
 \end{center}
\end{figure}
\vspace{3mm}

\item{}

\begin{figure}[h]
\begin{center}
\hspace*{-5mm}
\includegraphics[width=6cm]{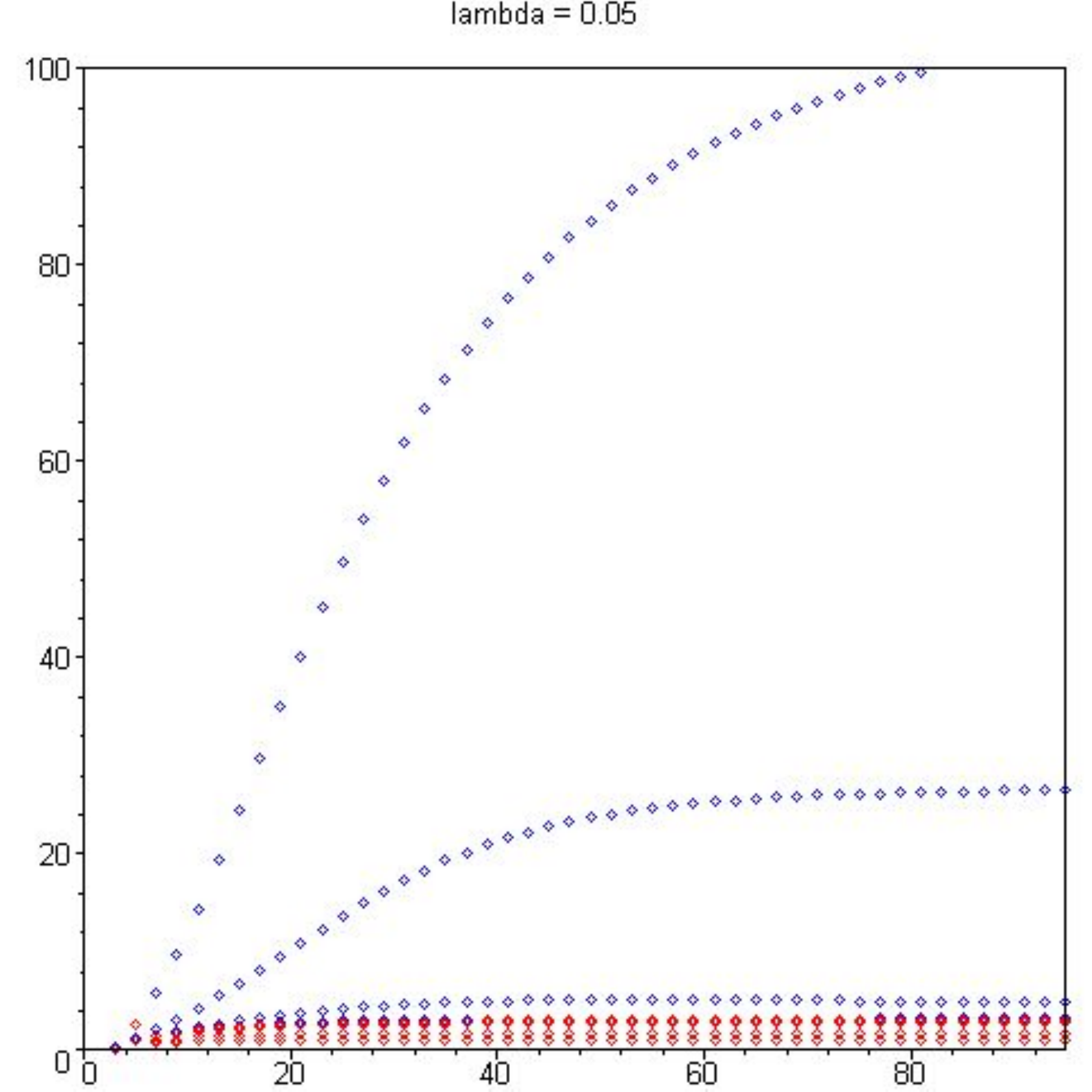}
 \caption{\underline{\textbf{$1^{rst}$ set:}}   Convergence up to $\nu= 6$ with
$\Lambda=0.05$  starting from $\delta_{n, max}$ and $\delta_{n, min}$}
 \label{fig.8}
 \end{center}
\end{figure}
\vspace{3mm}

\item{}

\begin{figure}[h]
\begin{center}
\hspace*{-5mm}
\includegraphics[width=6cm]{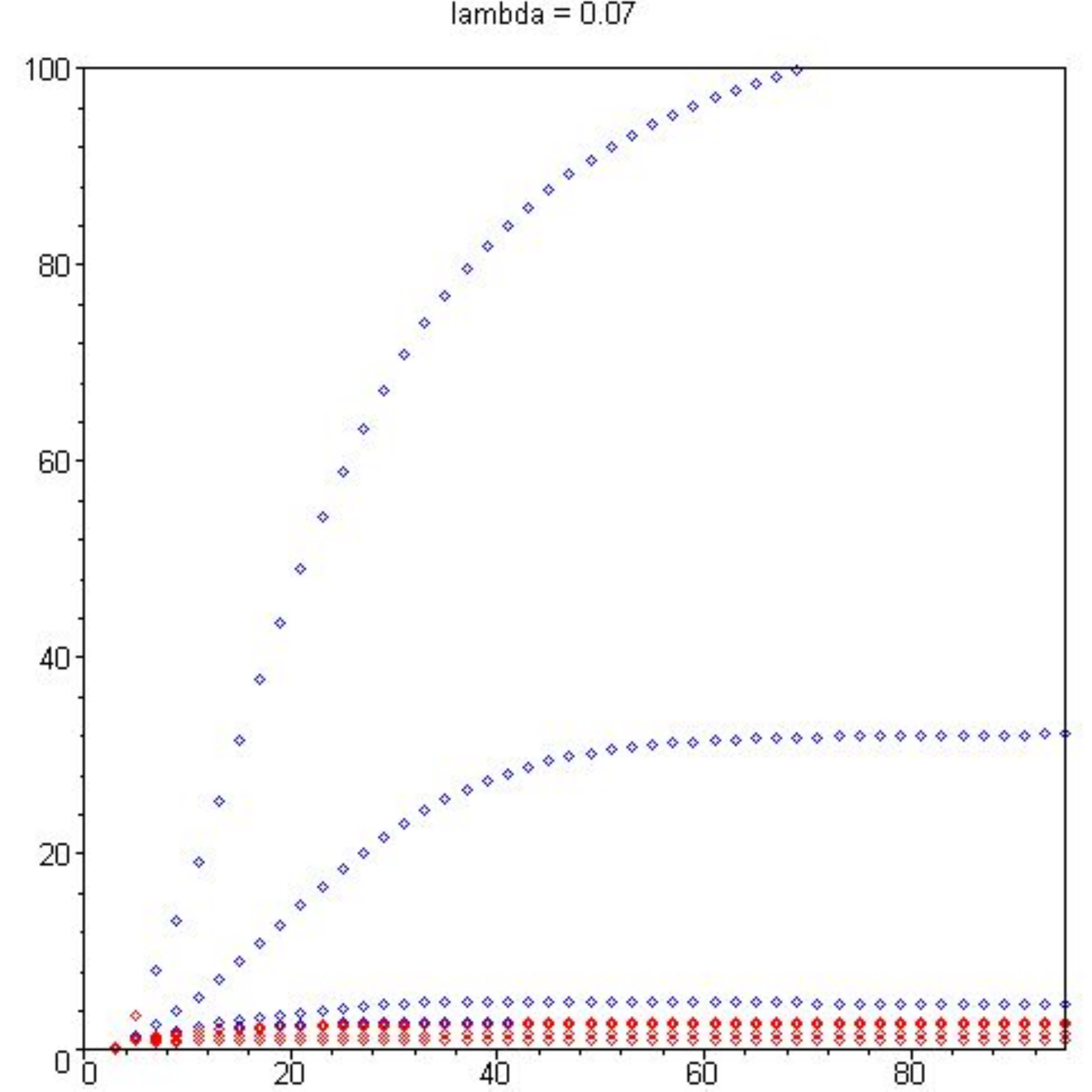}
 \caption{\underline{\textbf{$1^{rst}$ set:}}  Convergence up to $\nu= 6$ with
$\Lambda=0.07$  starting from $\delta_{n, max}$ and $\delta_{n, min}$}
 \label{fig.9}
 \end{center}
\end{figure}
\vspace{3mm}

\item{}

\begin{figure}[h]
\begin{center}
\hspace*{-5mm}
\includegraphics[width=6cm]{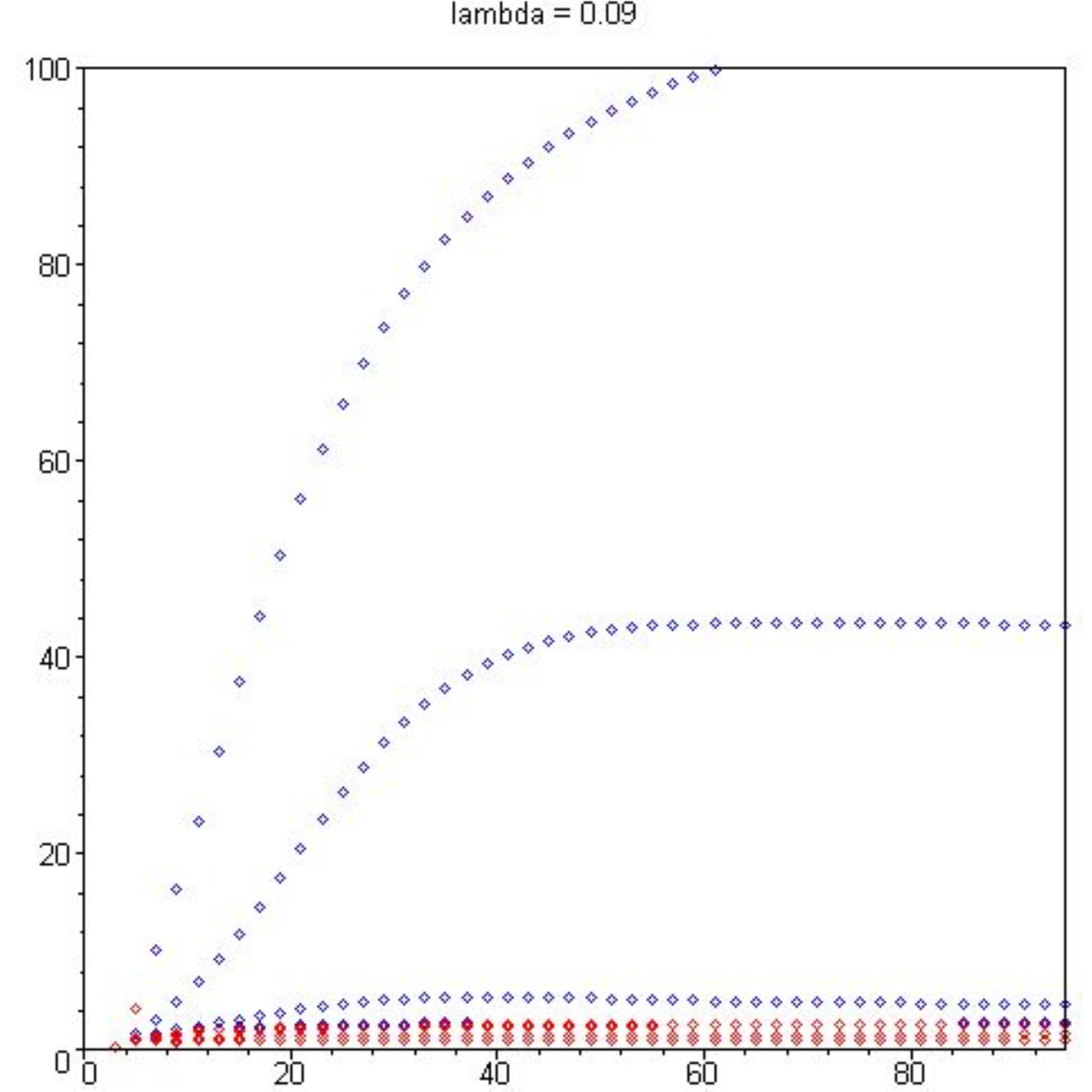}
 \caption{\underline{\textbf{$1^{rst}$ set:}}  Convergence up to $\nu= 6$ with
$\Lambda=0.09$  starting from $\delta_{n, max}$ and $\delta_{n, min}$}
 \label{fig.10}
 \end{center}
\end{figure}
\vspace{3mm}

\item{}

\begin{figure}[h]
\begin{center}
\hspace*{-5mm}
\includegraphics[width=6cm]{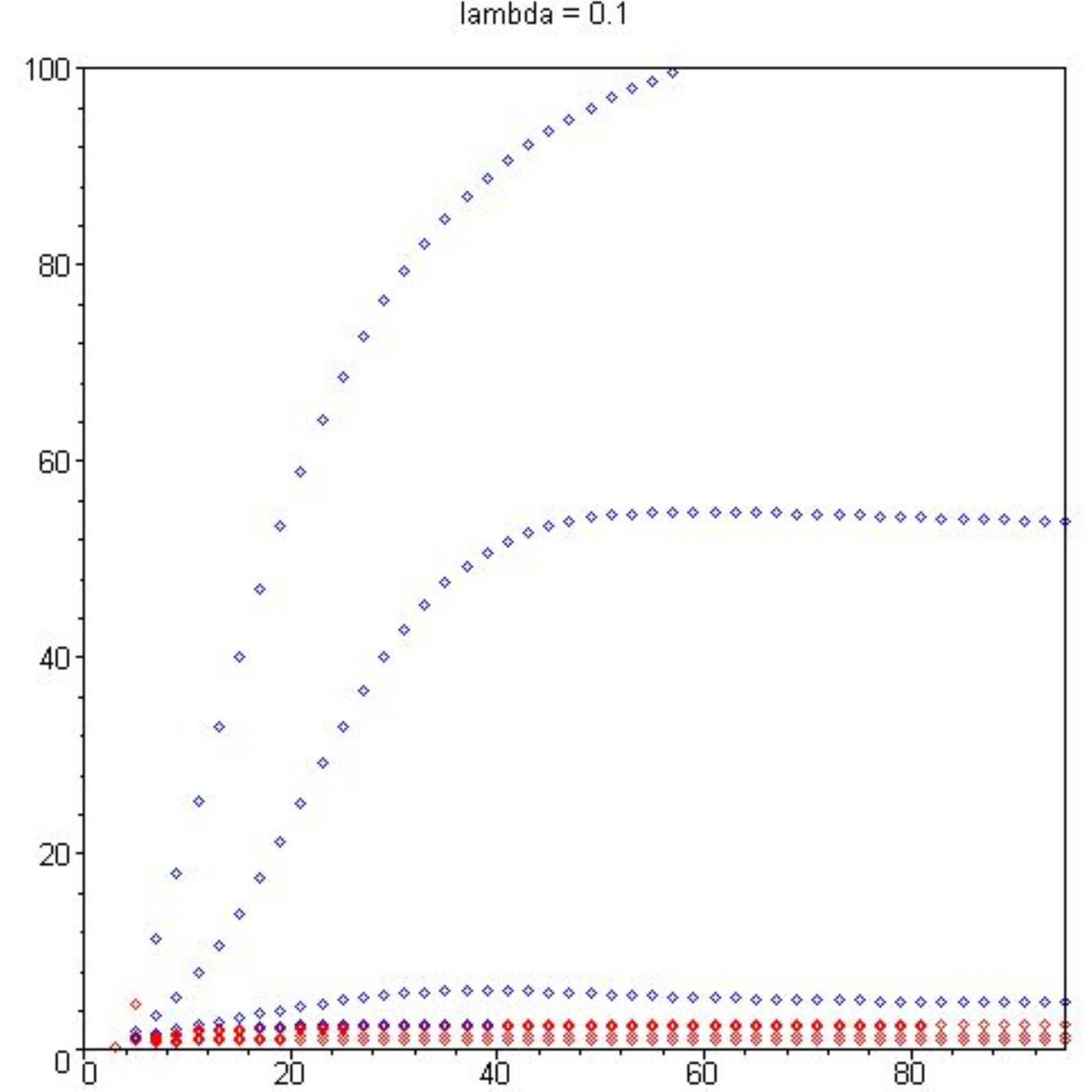}
 \caption{\underline{\textbf{$1^{rst}$ set:}} Convergence up to $\nu= 6$ with
$\Lambda=0.1$  starting from $\delta_{n, max}$ and $\delta_{n, min}$}
 \label{fig.11}
 \end{center}
\end{figure}
\vspace{3mm}

\item{}

\begin{figure}[h]
\begin{center}
\hspace*{-5mm}
\includegraphics[width=6cm]{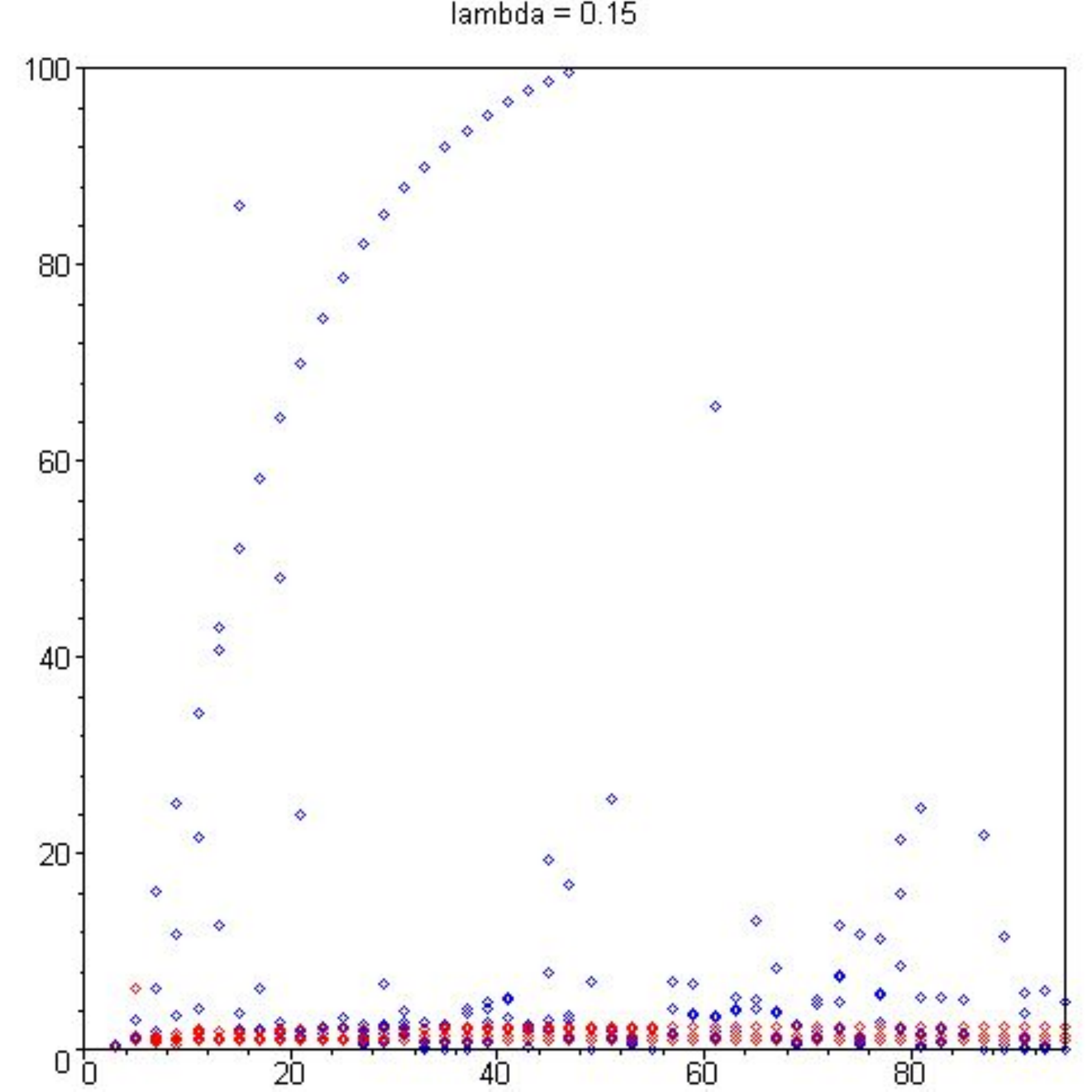}
 \caption{\underline{\textbf{$1^{rst}$ set:}} Divergence up to $\nu= 6$ with
$\Lambda=0.15$  starting from $\delta_{n, max}$ and $\delta_{n, min}$}
 \label{fig.12}
 \end{center}
\end{figure}

\vfill\eject

\item{}

\begin{figure}[h]
\begin{center}
\hspace*{-5mm}
\includegraphics[width=6cm]{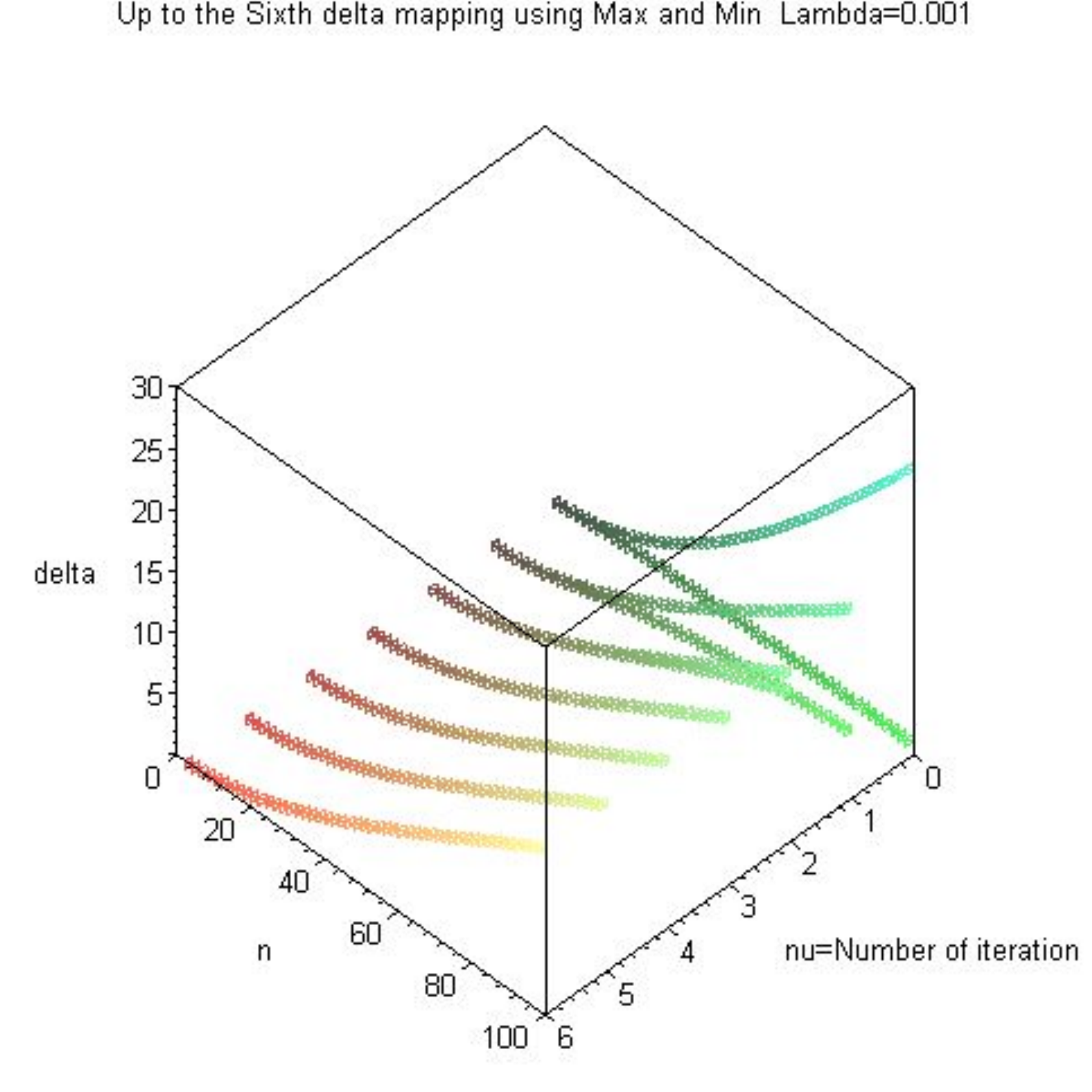}
 \caption{\underline{\textbf{$2^{nd}$ set:}}  Convergence up to $\nu= 6$ with
$\Lambda=0.001$  starting from $\delta_{n, max}$ and $\delta_{n, min}$}
 \label{fig.13}
 \end{center}
\end{figure}
\vspace{3mm}

\item{}

\begin{figure}[h]
\begin{center}
\hspace*{-5mm}
\includegraphics[width=6cm]{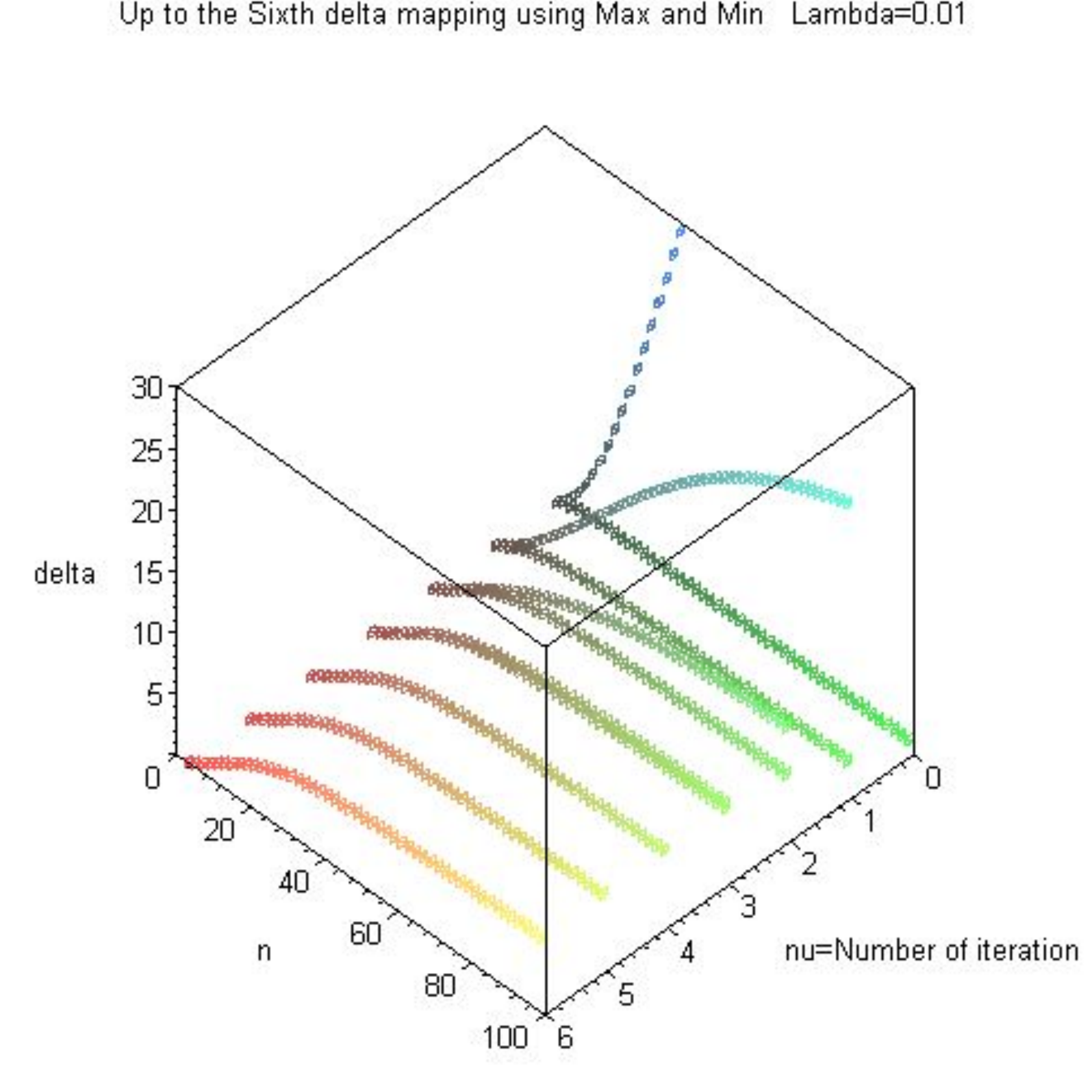}
 \caption{\underline{\textbf{$2^{nd}$ set:}}  Convergence up to $\nu= 6$ with
 $\Lambda=0.01$, starting from $\delta_{n, max}$ and $\delta_{n, min}$}
 \label{fig.14}
 \end{center}
\end{figure}

\item{}

\begin{figure}[h]
\begin{center}
\hspace*{-5mm}
\includegraphics[width=8cm]{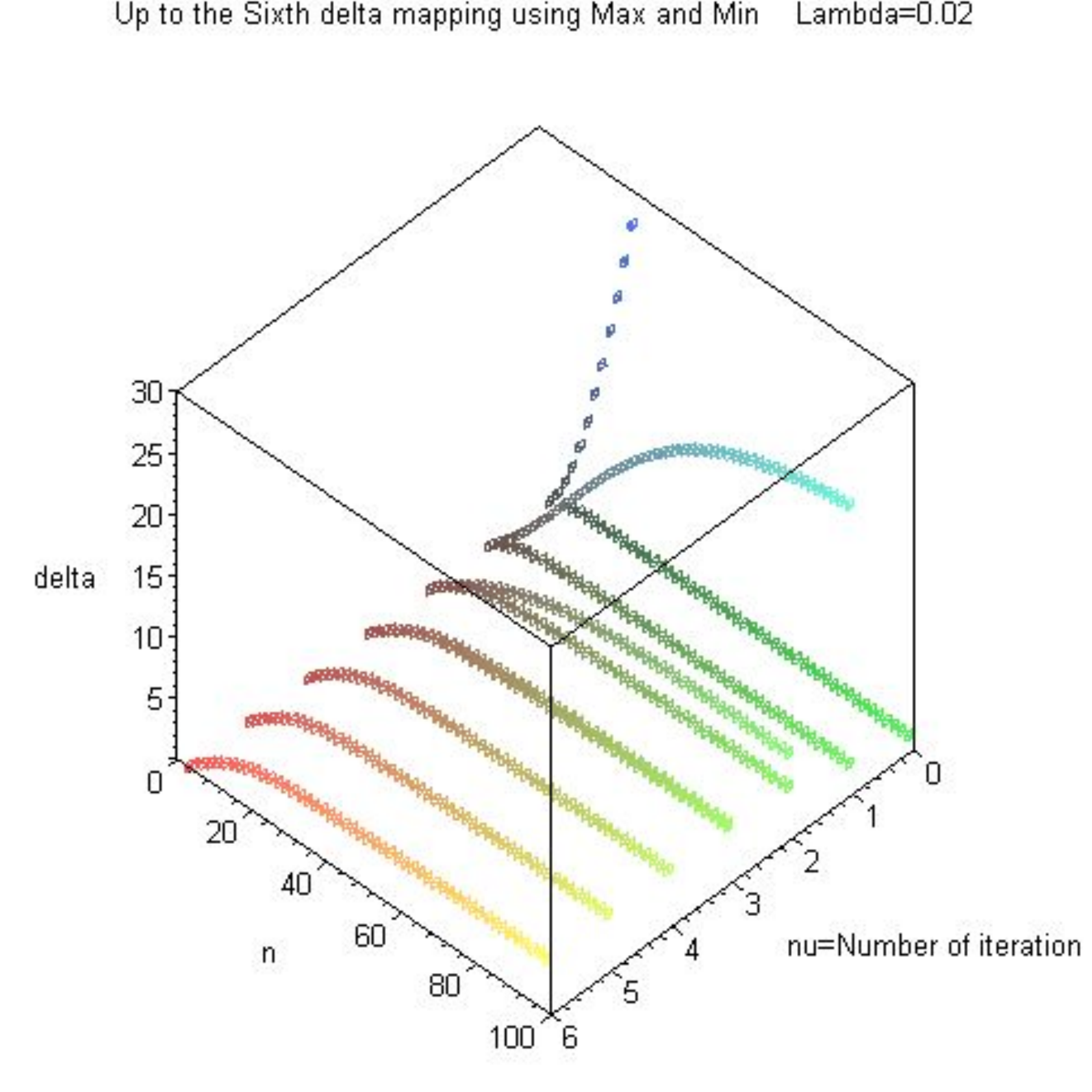}
 \caption{\underline{\textbf{$2^{nd}$ set:}}  Convergence up to $\nu= 6$ with
 $\Lambda=0.02$, starting from $\delta_{n, max}$ and $\delta_{n, min}$}
 \label{fig.15}
 \end{center}
\end{figure}

\item{}
\begin{figure}[h]
\begin{center}
\hspace*{-5mm}
\includegraphics[width=8cm]{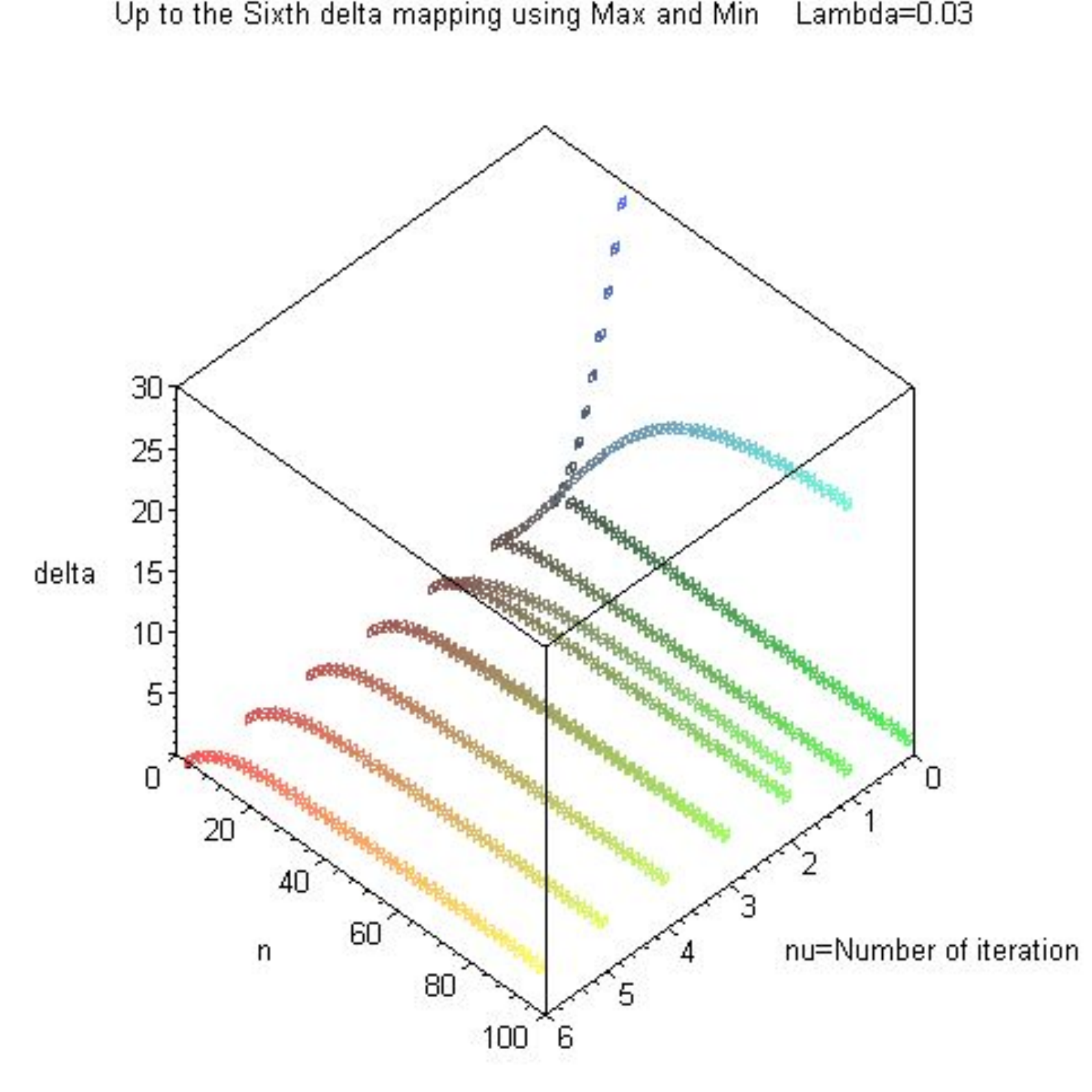}
 \caption{\underline{\textbf{$2^{nd}$ set:}}  Convergence up to $\nu= 6$ with
 $\Lambda=0.03$, starting from $\delta_{n, max}$ and $\delta_{n, min}$}
 \label{fig.16}
 \end{center}
\end{figure}

\item{}

\begin{figure}[h]
\begin{center}
\hspace*{-5mm}
\includegraphics[width=8cm]{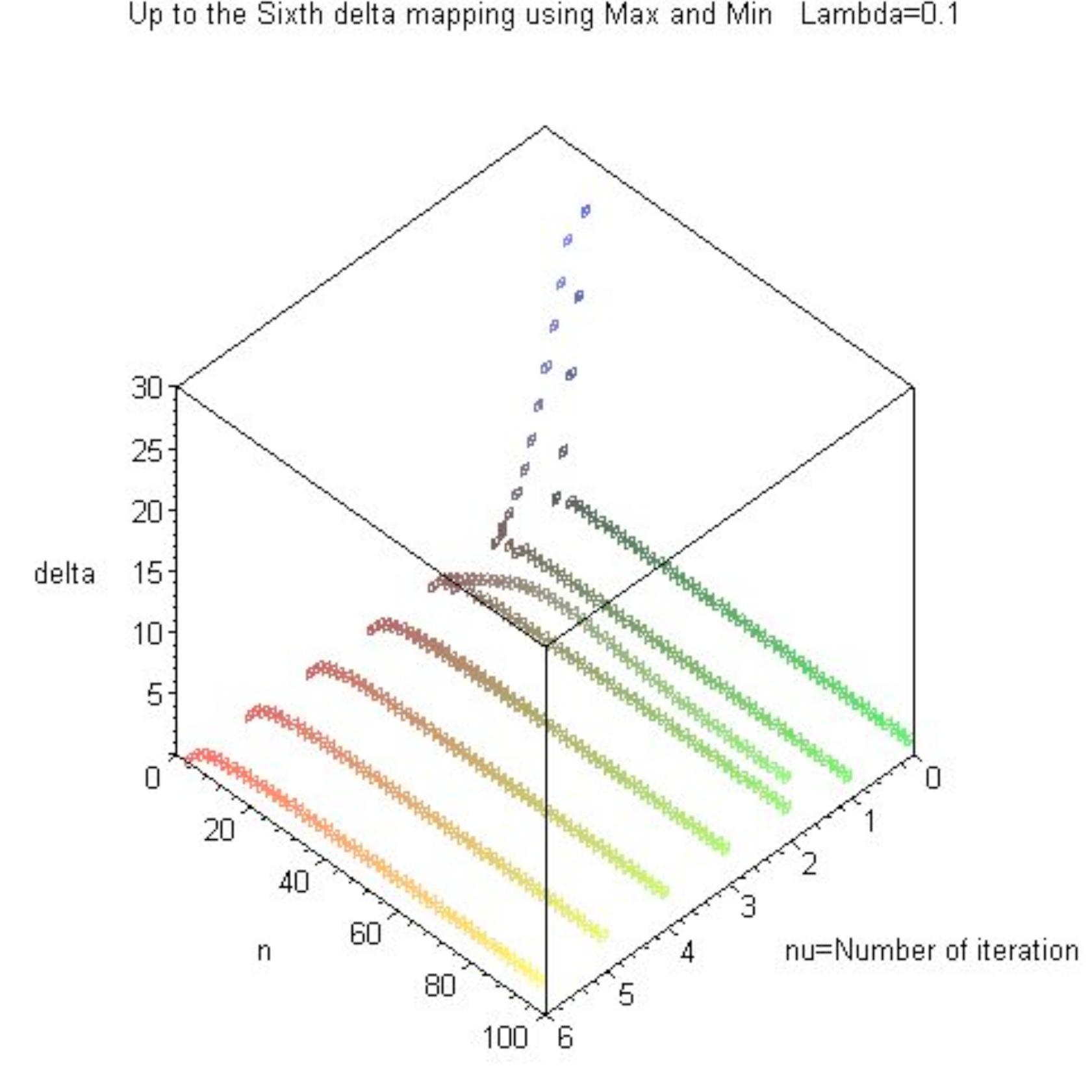}
 \caption{\underline{\textbf{$2^{nd}$ set:}}  Convergence up to $\nu= 6$ with
 $\Lambda=0.1$,  starting from  $\delta_{n, max}$ and $\delta_{n, min}$ }
 \label{fig.17}
 \end{center}
\end{figure}

\item{}

\begin{figure}[h]
\begin{center}
\hspace*{-5mm}
\includegraphics[width=8cm]{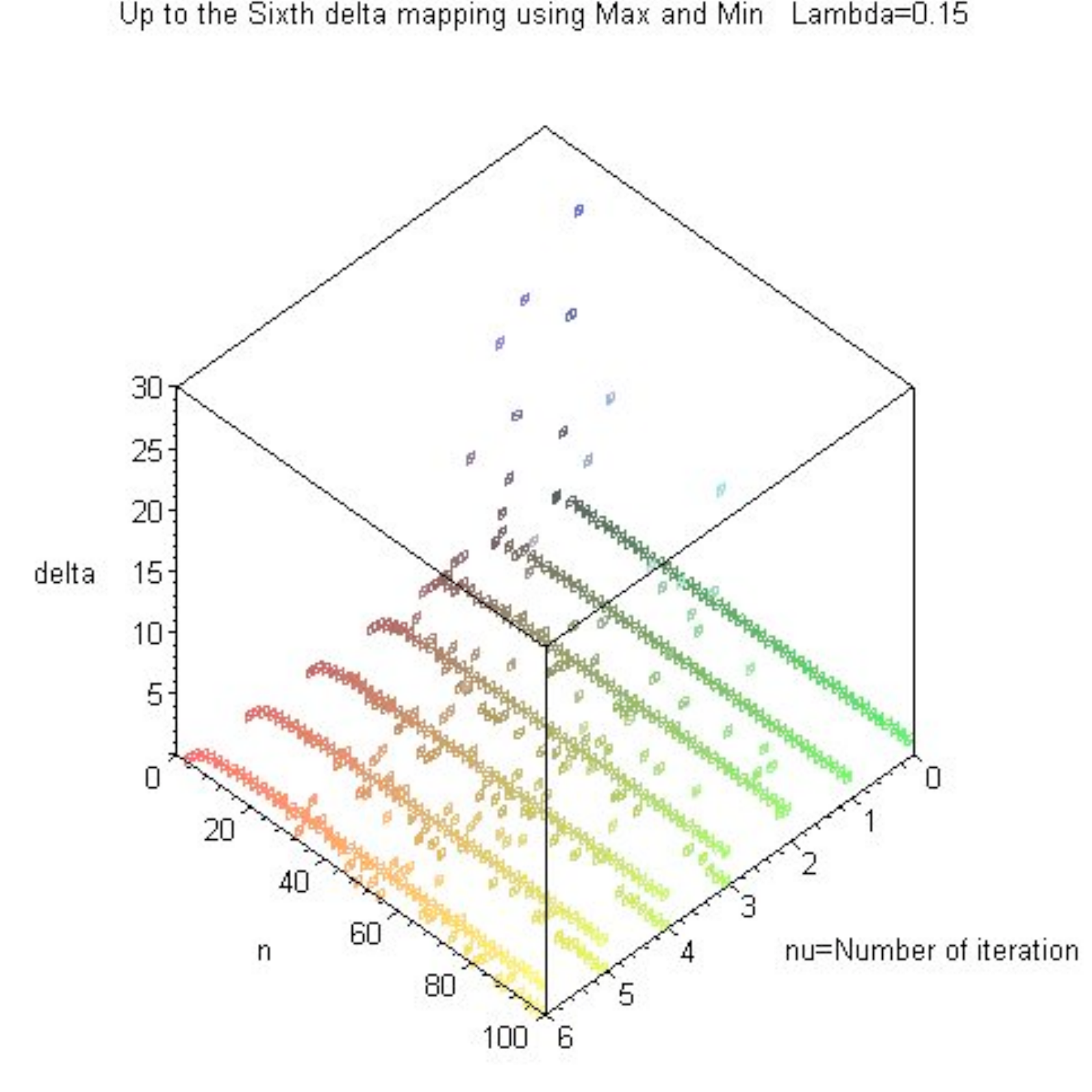}
 \caption{\underline{\textbf{$2^{nd}$ set:}} Divergence  (up to $\nu= 6$) with
 $\Lambda=0.15$, starting from $\delta_{n, max}$ and $\delta_{n, min}$}
 \label{fig.18}
 \end{center}
\end{figure}

\vfill\eject

\begin{figure}[h]
\begin{center}
\hspace*{-5mm}
\includegraphics[width=8cm]{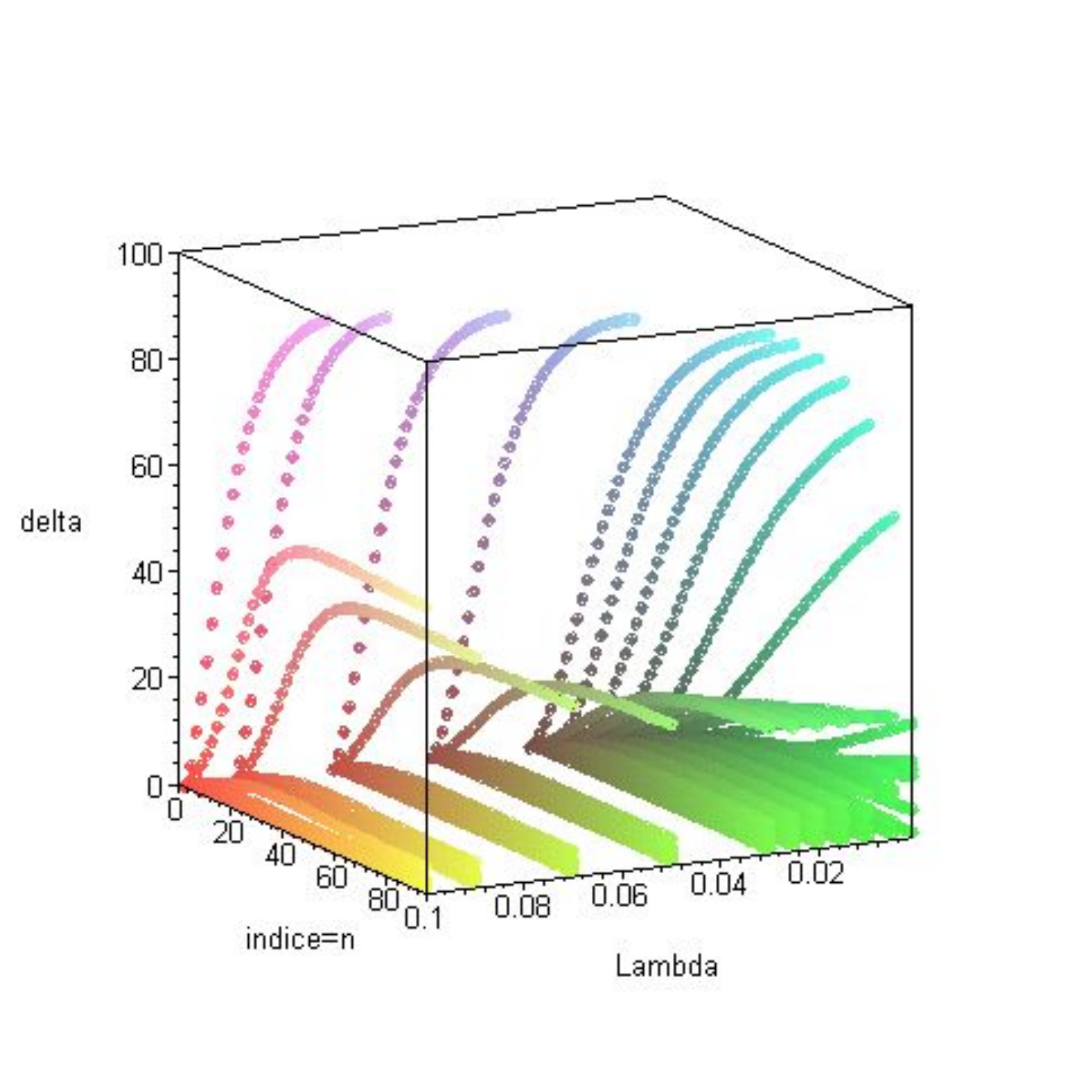}
 \caption{\underline{\textbf{$3^{d}$ set:}}  Summary up to the sixth iteration for
 $\Lambda$ from $0.001$ to $0.01$}
 \label{fig.19}
 \end{center}
\end{figure}

\begin{figure}[h]
\begin{center}
\hspace*{-5mm}
\includegraphics[width=8cm]{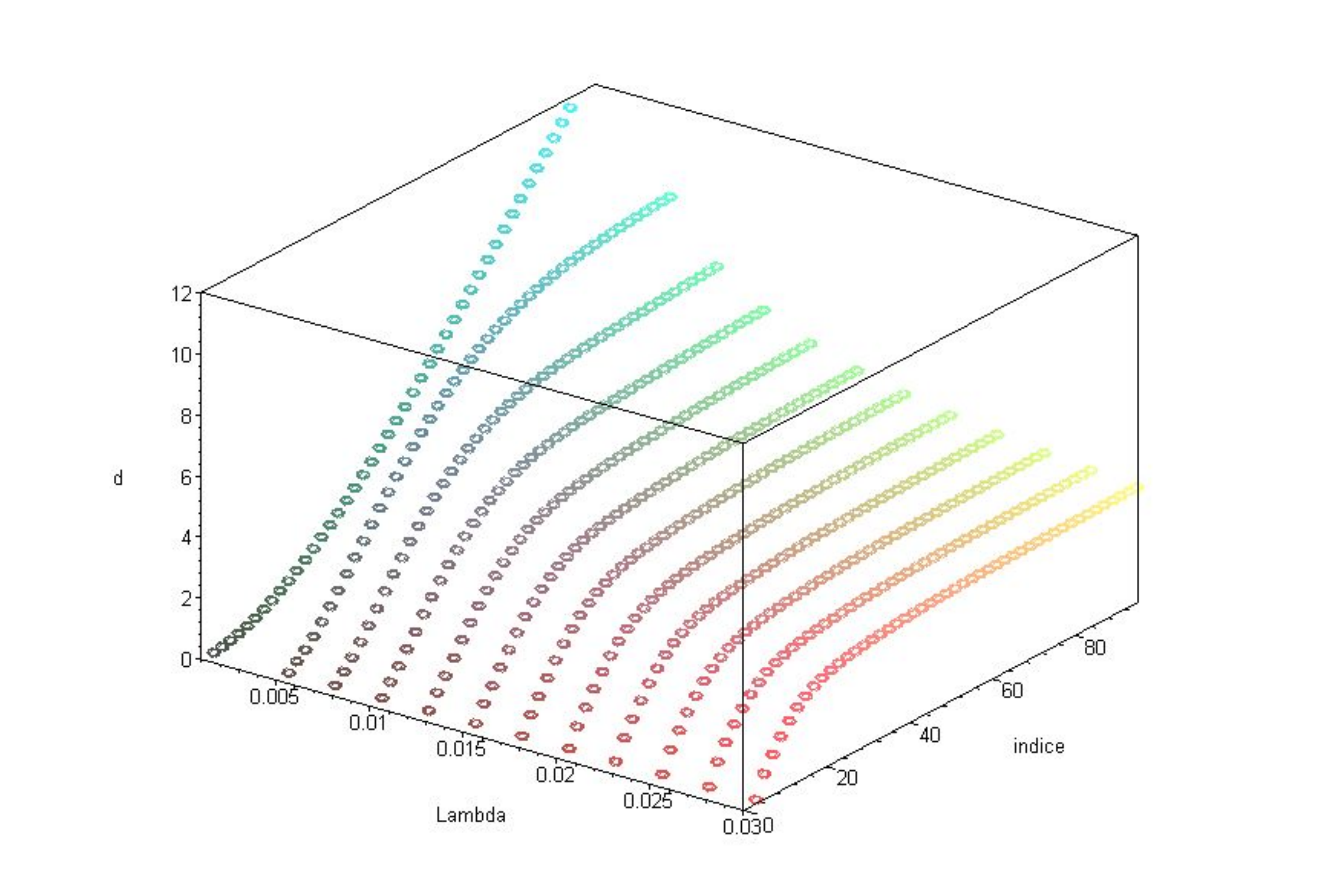}
 \caption{ \underline{\textbf{$3^{d}$ set:}} Summary of the six  iterations for
 $\Lambda$ from $0.001$ to $0.03$}
 \label{fig.20}
 \end{center}
\end{figure}

\end{enumerate}


\begin{thebibliography}{99}

\bibitem{(Q.F.T.)}
    \begin{itemize}
\item[a)] A.S. Wightman, Phys. Rev. 101, 860 (1965)

    \item[b)] R. Streater and A. Wightman.
\emph{PCT Spin Stat.and all That} (Benjamin, New York,1964)

    \item[c)] N.N. Bogoliubov, A.A. Logunov, and I.T. Todorov.
 \emph{Introduction to the Axiomatic Q.F.T.}
(Benjamin, New York, 1975)

        \item[d)] R. Jost. \emph{The General Theory
of Quantized Fields}
         (American Math.Society, Providence,RI,1965)
\end{itemize}
\bibitem{MM1}M. Manolessou
\begin{itemize}
\item[a)] J. Math. Phys. 20 (1988) 2092 
    \item   [b)] 30 (1989)175 
        \item [c)]  30 (1989) 907 
\item[d)] J. Math. Phys.32  (1991) 12 
    \item[e)] \emph{Back to the  $\Phi^4_0$  solution}
 Preprint  E.I.S.T.I. (1994)
 \item[{f)}] S. Gladkoff, A. Alaie,  Y. Sansonnet,  M. Manolessou
 J. Nonlin. Math. Phys. 9   2002, 77-85  (Electronic and Printed version)
\end{itemize}
    \bibitem{MM2} M. Manolessou
\begin{itemize}
\item[a)] Nucl. Physics B (Proc. Suppl.) 6 (1989) 163-166
North-Holland \item[b)]\emph{The  $\Phi^4_4$ non trivial solution}
 Preprint  E.I.S.T.I. (1992)
\item[c)] Contribution to the $XI^{th}$
                                        International Congress of Math.Physics Unesco-Sorbonne
                                        (D. Iagolnitzer editor 1994)
\end{itemize}
\bibitem{MM} M. Manolessou ``Local Contractivity of the $\Phi^4_0$ mapping'' Preprint EISTI September 2011
\end{thebibliography}
    \end{document}